# MP-PCA denoising of fMRI time-series data can lead to artificial activation "spreading"

Francisca F. Fernandes[a], Jonas L. Olesen[b,c], Sune N. Jespersen[b,c], Noam Shemesh[a,*]

[a]Champalimaud Research, Champalimaud Foundation, Lisbon, Portugal

[b]Center of Functionally Integrative Neuroscience (CFIN) and MINDLab, Department of Clinical Medicine, Aarhus University, Aarhus, Denmark

[c]Department of Physics and Astronomy, Aarhus University, Aarhus, Denmark

*Corresponding author:

Dr. Noam Shemesh, Champalimaud Research, Champalimaud Foundation, Av. Brasilia 1400-038, Lisbon, Portugal

E-mail: noam.shemesh@neuro.fchampalimaud.org;

Phone number: +351 210 480 000 ext. #4467

# Abstract

MP-PCA denoising has become the method of choice for denoising in MRI since it provides an objective threshold to separate the desired signal from unwanted thermal noise components. In rodents, thermal noise in the coils is an important source of noise that can reduce the accuracy of activation mapping in fMRI. Further confounding this problem, vendor data often contains zero-filling and other effects that may violate MP-PCA assumptions. Here, we develop an approach to denoise vendor data and assess activation "spreading" caused by MP-PCA denoising in rodent task-based fMRI data. Data was obtained from N = 3 mice using conventional multislice and ultrafast acquisitions (1 s and 50 ms temporal resolution, respectively), during visual stimulation. MP-PCA denoising produced SNR gains of 64% and 39% and Fourier spectral amplitude (FSA) increases in BOLD maps of 9% and 7% for multislice and ultrafast data, respectively, when using a small [2 2] denoising window. Larger windows provided higher SNR and FSA gains with increased spatial extent of activation that may or may not represent real activation. Simulations showed that MP-PCA denoising causes activation "spreading" with an increase in false positive rate and smoother functional maps due to local "bleeding" of principal components, and that the optimal denoising window for improved specificity of functional mapping, based on Dice score calculations, depends on the data's tSNR and functional CNR. This "spreading" effect applies also to another recently proposed low-rank denoising method (NORDIC). Our results bode well for dramatically enhancing spatial and/or temporal resolution in future fMRI work, while taking into account the sensitivity/specificity trade-offs of low-rank denoising methods.

# 1. Introduction

Noise is one of the main limiting factors in MRI, hindering the quality of multi-contrast data especially when high spatial or temporal resolutions are sought. Functional MRI (fMRI), in particular, can greatly suffer from noise: even with state-of-the-art equipment and tasks that elicit large activation, blood oxygenation level-dependent (BOLD) responses are inherently small (~0.5-3%), leading to a low contrast-to-noise ratio (CNR). Moreover, ultrafast acquisitions in rodents are emerging as a novel means for better characterization of BOLD responses and for better representation of neural information flow[1-4]. Pre-clinical fMRI in rodents is also important for multimodal studies that can provide excellent genetic flexibility and unique experimental settings such as optogenetic control of specific neurons[5], simultaneous calcium recordings[6] and imaging[7], and multitude models of disease[8]. However, the demand for higher spatiotemporal resolution comes with a decrease in the signal-to-noise ratio (SNR), which hinders the accuracy and precision of functional activation estimation. Even though this can be in part ameliorated by using cryogenic coils[3,9,10], data denoising in pre-clinical fMRI remains critical.

Different denoising approaches for fMRI have been developed throughout the years. Total variation minimization[11,12] and non-local means[13-16] have proven useful to reduce noise in different MRI applications. For fMRI in particular, spatial and temporal filtering[17-19] and the addition of noise regressors to general linear model (GLM) fitting (GLMdenoise)[20] have enabled improved activation mapping. However, despite the usefulness of these techniques, they are either limited by a loss in spatial resolution (smoothing, blurring of anatomical details and introduction of partial volume effects) that leads to compromised accuracy in further quantitative analyses, or the need of subjective user input.

Principal component analysis (PCA)-based approaches[21-24] have been proposed to address these issues. Most of the signal-related variance of redundant data is contained within just a few principal components while the noise is evenly spread over all components, allowing the reduction of thermal noise without requiring training data or accurate brain region segmentation. Although these approaches have exhibited very good potential for increasing BOLD CNR in fMRI data[25,26], they require a subjective and often empirically defined[22] user-input threshold for determining which principal components to keep, and which to reject as "noise". The lack of objective methods for defining these thresholds can lead to inconsistency between sites and, if the wrong threshold is applied, could also risk the removal of important information.

To overcome this limitation, Marchenko-Pastur PCA (MP-PCA) denoising was recently proposed[27,28]. Using random matrix theory principles developed early on by Marchenko and Pastur[29], MP-PCA provides an objective data-driven (model-independent) threshold to identify noise components based on asymptotic mathematical properties of the eigenspectrum of random covariance matrices. In the large matrix size limit, the noise contribution to the histogram of PCA eigenvalues universally follows the MP distribution[29]. Therefore, by leveraging MRI data redundancy, MP-PCA denoising is able to estimate the local noise level and target its suppression in a deterministic way. Moreover, MP-PCA is a selective denoising technique, as it removes – ideally – only thermal noise and not any other structured non-white noise that arises from respiratory, cardiac or spontaneous neuronal activity (physiological noise), motion or artifacts, thereby preserving fine anatomical detail. Currently, MP-PCA is the most widespread method for noise suppression in diffusion MRI (dMRI)[28,30-32] and it has also been recently successfully applied to MRI relaxometry data[32,33].

Given that fMRI data are typically redundant, MP-PCA denoising was recently also proposed for fMRI. Particularly, improvements were shown for task-based fMRI language mapping in humans[34] and enhanced temporal SNR (tSNR) and network mapping in human and rat resting-state fMRI data [35,36]. However, the effects of MP-PCA denoising on spatial specificity of functional maps have never been comprehensively assessed. Moreover, MP-PCA has not yet been explored in pre-clinical task-based fMRI, which might be a missed opportunity since rodent fMRI suffers mostly from random signal fluctuations caused by thermal noise in the smaller coils.

Despite the proven ability of MP-PCA to significantly improve the quality of clinical and scientific MRI data, there are some aspects that limit the application of this technique. In the practical finite matrix case, deviations from the asymptotic expression may occur in MP-PCA analyses and "tails" may appear near the right edge of the MP distribution. As a way to solve this problem, the NOise Reduction with DIstribution Corrected (NORDIC) PCA approach was recently developed, initially for dMRI [37] and more recently for fMRI [38], showing improvements in key metrics of functional mapping while avoiding image blurring or decrease of spatial precision. Like the MP-PCA denoising method, NORDIC is a low-rank patch-based denoising method that works directly on the spectrum of the Casorati matrix with the sole purpose of suppressing thermal noise. However, NORDIC numerically estimates the threshold via a Monte Carlo simulation with random noise finite matrices, whose variance matches experimentally measured thermal noise, to generate the sample average for the largest PCA eigenvalue. Moreover, NORDIC performs signal and noise scaling to ensure zero-mean and spatially identical noise prior to denoising. A further complication can arise from images reconstructed by the vendor, which may not be ideally suited for the immediate application of MP-PCA denoising. In particular, partial Fourier reconstructions[39] and acquisition along (e.g., EPI) gradient ramps, commonly used in fMRI to achieve faster acquisition rates, introduce zero-filling (ZF) in the phase encoding and readout directions of k-space, respectively, which generate spatial correlations that change the noise statistics, and violate the i.i.d. noise characteristics on which MP-PCA assumptions rely. Lastly, the projection to a local low-rank approximation of the data upon MP-PCA denoising may involve local "bleeding" of active signal components that result in "spreading" of fMRI activation patterns.

Here, we develop an approach to denoise pre-clinical vendor data and assess activation "spreading" caused by MP-PCA denoising in pre-clinical fMRI acquisitions. To that end, we acquired conventional multislice "slow" and more challenging ultrafast (temporal resolution = 50 ms) fMRI data from the mouse brain while performing a visual stimulation task known to elicit strong neuronal and BOLD responses[40-45]. Moreover, we use a more permissive MP-PCA threshold (see Methods) than the one described in the original version of the algorithm (Veraart et al.[28]), that takes into account the tails at the edge of the MP distribution arising from the finite matrix sizes, thus enabling the preservation of more signal components. We validate the performance using simulations of data mimicking fMRI signals with a known ground truth (GT), and finally, we extend our conclusions to the more recent NORDIC PCA denoising approach. Our findings identify denoising-driven artifactual activation spreading and provide means to avoid it based on SNR and CNR, which can be important for future studies seeking to harness denoising in fMRI.

## 2. Methods

All animal experiments were preapproved by the institutional and national authorities and carried out in accordance to European Directive 2010/63.

2.1. Animal preparation

Adult C57BL/6 male mice (N = 3) weighing 24.9 ± 1.0 g and aged 8.0 ± 0.3 weeks were used in this study (animal weights and ages are reported as mean ± SD). Animals were reared in a temperature-controlled room and held under a 12h/12h light dark regime with *ad libitum* access to food and water. Anesthesia was induced with 5% isoflurane (Vetflurane®, Virbac, France) mixed with oxygen-enriched (28%) medical air. Animals were then weighed, moved to the animal bed (Bruker BioSpin, Germany) and isoflurane was reduced to 2-3%. Ear and bite bars were fixed and eye drops (Bepanthen® Eye Drops, Bayer AG, Germany) were applied to prevent eye dryness. A 0.4 mg/kg subcutaneous bolus of medetomidine (Dormilan®, Vetpharma Animal Health S.L., Spain; 1 mg/ml, diluted 1:10 in saline) was injected ~6.5 min after isoflurane induction, followed by a gradual discontinuation of isoflurane to 0% over the next 10 min. After this time period, a continuous infusion of medetomidine (0.8 mg/kg/h) was initiated and maintained until the end of the experiment[46].

Breathing rate and rectal temperature were monitored throughout the sessions using a pillow sensor and an optic fiber probe (SA Instruments, Inc., USA), respectively. A warm-water recirculating pad was used for body temperature control (maintaining temperatures between 35.0-36.8 °C). At the end of experiments, a subcutaneous injection of atipamezole (Antidorm, Vetpharma Animal Health S.L., Spain; 5 mg/ml, diluted 1:10 in saline) at 2 mg/kg was given to all animals to reverse the medetomidine effects.

2.2. MRI protocol

Animals were imaged in a 9.4 T BioSpec MRI scanner (Bruker, Karlsruhe, Germany) equipped with an AVANCE III HD console, producing isotropic pulsed field gradients of up to 660 mT/m (120 μs rise time), an 86 mm-ID quadrature resonator for RF transmission and a 10 mm loop surface coil for signal reception, thus avoiding coil-driven spatial correlation of the noise due to multiple-channel receiver coil (e.g. cryoprobe) combinations. The scanner runs ParaVision 6.0.1 software (Bruker, Ettlingen, Germany).

Following routine adjustments for center frequency, RF calibration, acquisition of $B_0$ maps and automatic shimming, anatomical images were acquired using a $T_2$-weighted Turbo

RARE sequence in the coronal, sagittal and oblique planes (TR/TE = 2000/40 ms, FOV = 20×16 mm$^2$, in-plane resolution = 80×80 μm$^2$, RARE factor = 5, slice thickness = 0.5 mm, number of slices = 20 (if coronal or oblique) or 23 (if sagittal), $t_{acq}$ = 1 min 18 s). An extra acquisition with 3 oblique slices (TR/TE = 2000/40 ms, FOV = 16×12.35 mm$^2$, in-plane resolution = 64×62 μm$^2$, RARE factor = 5, slice thickness = 1 mm, $t_{acq}$ = 1 min 18 s) was also performed for later coregistration of ultrafast fMRI scans.

Functional imaging began ~40 min after isoflurane induction[47]. Conventional multislice fMRI acquisitions were performed using a GE-EPI sequence: TR/TE = 1000/15 ms, flip angle = 60°, FOV = 16×12 mm$^2$, in-plane resolution = 145×145 μm$^2$, partial Fourier factor in the PE direction = 0.8, slice thickness = 0.5 mm, number of slices = 10, repetitions = 340, $t_{acq}$ = 5 min 40 s. Ultrafast fMRI acquisitions with a single oblique slice capturing the entire visual pathway[3] (GE-EPI, TR/TE = 50/17.5 ms, flip angle = 15°, FOV = 16×12.35 mm$^2$, resolution = 167×167 μm$^2$, partial Fourier factor along the PE direction = 0.8, slice thickness = 1 mm, repetitions = 9000, $t_{acq}$ = 7 min 30 s, dummy scans = 5240 to ensure the coil reaches a steady temperature during these intensive acquisitions) were also performed.

2.3. Visual stimulation

As depicted in Figure 1A, left eye monocular flashing stimulation (frequency = 2 Hz; pulse width = 10 ms) was performed using a blue (wavelength = 470 nm) LED connected to an optical fiber that extended to the interior of the scanner (at ~1 cm from the eye). The right eye was covered with a patch. The stimulation paradigms consisted of five blocks of 20 s stimulation and 40 s rest for multislice fMRI, and ten cycles of 1 s stimulation and 40 s rest for ultrafast fMRI. Both acquisitions were alternated and repeated 2-4 times per animal, and separated by a resting period of at least 6 min to avoid habituation.

2.4. Data analysis

All datasets were analyzed in MATLAB (MathWorks, USA).

Data was first corrected for outliers by manually selecting time points whose average brain signal strongly deviated (± 2-3 SD) from its 2$^{nd}$ order polynomial trend, and estimating new voxel values at those time points using piecewise cubic interpolation from the signal at the remaining time points. This correction was performed independently for each slice. Only < 0.2% of datapoints/scan were corrected in multislice fMRI data, whereas no outliers were found in ultrafast data.

As shown in the diagram of Figure 1B, after outlier correction, data was denoised using two different strategies. In Strategy A, data was MP-PCA denoised in its magnitude form using 15 different sliding windows in case of multislice fMRI (size varying between [2 2], [5 5], [10 10], [20 20] and [40 40] in the row and columns dimensions, and between 1, 5 and 10 in the slice dimension) and 6 in case of ultrafast fMRI ([2 2], [5 5], [20 20], [40 40], [60 60] and [74 96], where the last one corresponds to the entire image matrix). In Strategy B, data was first transformed in its complex form into k-space for ZF removal. After returning to image space, complex images were MP-PCA denoised at the slice level using 8 different sliding windows ([2 2], [3 3], [4 4], [5 5], [10 10], [15 15], [20 20] and [25 25]), and then converted to magnitude images. Data patches used for denoising were maximally overlapping (i.e., the window slid voxel-by-voxel) and were equally averaged after denoising. Moreover, we used a slightly improved version of the algorithm suggested by Veraart et al.[28] that has shown better results in the estimation of the number of significant signal components (https://github.com/sunenj/MP-PCA-Denoising)[48]. Particularly, instead of increasing the number of signal components $p$ from 0 until

$$\frac{1}{(M'-p)N'}\sum_{i=p+1}^{M'} \lambda_i \geq \frac{\lambda_{p+1}-\lambda_{M'}}{4}\frac{1}{\sqrt{N'(M'-p)}} \quad (1)$$

we used

$$\frac{1}{(M'-p)(N'-p)}\sum_{i=p+1}^{M'} \lambda_i \geq \frac{\lambda_{p+1}-\lambda_{M'}}{4}\frac{1}{\sqrt{N'M'}} \quad (2)$$

where $M$ is the number of image repetitions, $N$ is the total number of voxels within the sliding window, $\lambda$ are the PCA eigenvalues, $M' = \min(M,N)$ and $N' = \max(M,N)$.

Further pre-processing included slice-timing correction (only for multislice data), motion correction, coregistration, normalization to the Allen Reference Atlas[49] and smoothing with a 3D isotropic Gaussian kernel with FWHM = 145 μm (if multislice fMRI) or 167 μm (if ultrafast fMRI). These steps were performed using SPM12 tools.

Image residuals were calculated after MP-PCA denoising to evaluate the algorithm performance, by subtracting the original undenoised images from the denoised images. The residuals were then divided by the map of estimated noise standard deviation $\sigma$ (which is an output of the MP-PCA denoising procedure, obtained by voxel-wise averaging of the $\sigma$ values estimated for each patch). The normality of residuals was assessed by histogram analysis and normal distribution fitting. Once normality was confirmed, the variance of this distribution fit then allowed to compute the percentage of estimated noise variance $\sigma^2$ explained by the residuals (equivalent to $\sigma^2_{residuals}/\sigma^2 \times 100$). The SNR gain obtained by the denoising was also calculated at this stage, by using the expected variance reduction $\sigma_r^2$ obtained by truncating Gaussian noise components only (i.e., $\sigma_r^2 = \frac{(M'-P)(N'-P)}{M'N'}\sigma^2$), and computing

$$\text{SNR}_{\text{gain}}(\%) = \left(\sqrt{\frac{\sigma^2}{\sigma^2 - \sigma_r^2}} - 1\right) \times 100 = \left(\sqrt{\frac{M'N'}{M'N' - (M'-P)(N'-P)}} - 1\right) \times 100 \quad (3)$$

where $P$ is the number of retained "signal" components given as an output of the method (obtained after voxel-wise averaging of the $P$ values estimated for each patch).

To avoid hemodynamic response function (HRF) and GLM-related a-priori assumptions, a data-driven Fourier approach was performed for BOLD mapping[3,50], where the Fourier Spectral Amplitude (FSA) at the paradigm's fundamental frequency (i.e., the first harmonic) was mapped voxelwise to detect activated areas. Single-voxel time-courses were detrended and standardized to z-score prior to fast Fourier transform computation. In case of ultrafast data, FSA values at the two following harmonics of the paradigm (i.e., the second and third harmonics) were also extracted and summed with the FSA at the fundamental frequency before mapping. Phase information was used to separate positive from negative responses. Specifically, the phase difference $\Delta\Phi$ between the data and the paradigm was computed and averaged at the chosen harmonics for each voxel. Values of $|\Delta\Phi| < \pi/2$ translated into positive responses, whereas $\pi/2 < |\Delta\Phi| < \pi$ translated into negative responses. Individual maps were then thresholded with an FSA threshold of 0.3 (multislice fMRI) or 0.08 (ultrafast fMRI), whereas group maps (computed from the voxel-timecourses averaged between a total of n = 9 multislice or ultrafast scans) were thresholded with an FSA threshold of 0.45 (multislice fMRI) or 0.09 (ultrafast fMRI). A minimum cluster size of 10 (multislice fMRI) or 8 (ultrafast fMRI) voxels was also imposed, based on the minimum size required to detect activation in the smallest regions of the visual pathway (i.e., the lateral posterior and geniculate nuclei of the thalamus). To quantify improvements from MP-PCA denoising on BOLD mapping, the average FSA in activated voxels (i.e., the voxels above the threshold in the BOLD map obtained from undenoised data) was extracted from each map and percentage increases relative to the undenoised data results were quantified. The number of activated voxels in each map was also calculated to obtain the increase in spatial extent of activation upon denoising. To understand if different functional CNR and temporal SNR (tSNR) values could explain the variations in the increase of FSA values between individual datasets, we run a Pearson's correlation to assess the relationship between those variables for each sliding window. The average FSA extracted from activated voxels of undenoised data BOLD maps was used as an indicator for functional CNR. The tSNR values were calculated from a resting period in the middle of the undenoised data acquisitions.

2.5. Simulations with MP-PCA denoising

To more deeply investigate changes in spatial extent of activation after denoising, we simulated data to mimic the ultrafast fMRI data but with an a-priori known GT. A double-gamma HRF was generated for every voxel of four different bilateral ROIs of the visual pathway (primary visual cortex (V1), superior colliculi (SC), lateral posterior (LP) and lateral geniculate (LGN) nuclei of the thalamus), with slight changes in latency and amplitude per ROI and voxel, and convolved with the ultrafast fMRI paradigm to simulate BOLD responses. In particular, the BOLD responses were built so that they would peak randomly within the [1.7,2.3] s interval after stimulus onset in voxels of the left hemisphere and within the [3.7,4.3] s interval in voxels of the right hemisphere. Moreover, the percent signal change of the BOLD signal was randomly distributed between 0.75-1.25% in V1 and LP and 1.75-2.25% in the SC and LGN. The temporal resolution was set to 50 ms. Gaussian white noise was then added to GT data so that the average tSNR in the brain = 6.7, i.e., similar to the average values observed in real data.

Simulated complex data were MP-PCA denoised using 8 different sliding windows ([2 2], [3 3], [4 4], [5 5], [10 10], [15 15], [20 20] and [25 25]) and maximally-overlapping patches. Residual maps were again calculated by subtracting the undenoised (noisy) images from the denoising images and dividing the result by the map of estimated noise $\sigma$. Error maps were generated by subtracting the noise-free GT data from undenoised or denoised images and dividing the result by the standard deviation $\sigma_s$ of the added noise. To characterize the strength and the pattern of the temporal correlation of neighboring voxels in the brain upon denoising, the noise kernel was estimated for each denoising window using LayNii[51].

BOLD Fourier maps were calculated and thresholded in the same way as for real data. Percentage increases of the average FSA in GT activated voxels relative to the undenoised data results were quantified. The increase in spatial extent of activation relative to the GT data was also calculated. To further quantify the performance of MP-PCA denoising in terms of spatial extent of activation, the Dice similarity coefficient was calculated. Moreover, the sensitivity and false positive rate (FPR) were quantitatively evaluated as in the study of Fang et al. [52]. Specifically, the sensitivity was quantified as the number of true positive voxels over the number of true positive and false negative voxels, and the FPR as the ratio between the number of false positive voxels and the number of false positive plus true negative voxels within the first to fifth pixel perimeter layers of the GT activation volume inside the brain mask ($FPR_1$ to $FPR_5$).

To study the performance of this denoising method in more extreme conditions, four supplementary simulations were also performed: two with more (tSNR = 3) or less (tSNR = 15) noise, and two with smaller or higher values of BOLD percent signal change, respectively, one with 0.10-0.50% changes in V1 and LP and 0.30-0.70% in the SC and LGN, and another with 2.75-3.25% changes in V1 and LP and 3.75-4.25% in the SC and LGN.

2.6. Simulations with NORDIC PCA denoising

To assess if NORDIC PCA denoising also causes activation "spreading" in BOLD maps, the first simulated ultrafast fMRI dataset (with tSNR = 6.7) was denoised with the NORDIC PCA approach using 5 different sliding windows ([2 2], [5 5], [10 10], [25 25] and [40 40]), maximally overlapping patches, and with and without the phase-stabilization correction approach (low-pass filter width = 10) described by Moeller et al.[37]. BOLD Fourier maps were calculated and thresholded as in the previous simulations.

2.7. General linear model analysis of functional data

Given that the majority of BOLD fMRI publications employ general linear model (GLM) for data analysis, we computed BOLD maps generated by GLM analysis for the acquired multislice fMRI datasets and the ultrafast fMRI simulation with tSNR = 6.7 and 0.75-2.25% BOLD changes to test the generalizability of our results. For both types of data, the experimental regressor of the design matrix was generated by convolution of a double gamma HRF with the respective paradigm, so that the expected BOLD response would peak at 3 s after stimulus onset. Motion correction parameters were used as nuisance regressors for multislice fMRI data. Resulting *t*-value maps were thresholded with a minimum *t*-value of 1.65 (equivalent to $p < 0.05$ in undenoised data) and a minimum cluster size of 10 (multislice fMRI) or 8 (ultrafast fMRI simulation) voxels. A fixed-effect group analysis was also run using the n = 9 multislice fMRI datasets. In this case, the resulting *t*-value maps were thresholded with a minimum *t*-value of 3.09 (equivalent to $p < 0.001$ in undenoised data) and a minimum cluster size of 10 voxels.

# 3. Results

3.1. Conventional multislice and ultrafast fMRI provide artifact-free images and robust activation maps in the mouse visual pathway.

Before application of MP-PCA denoising, we assessed the quality of the acquired multislice and ultrafast datasets. Figures 2A and 3A show a single GE-EPI image obtained from a representative multislice and ultrafast fMRI acquisition, respectively, revealing good brain contrast and no remarkable artifacts in both cases. The temporal SNR in the brain was 13.1 ± 0.7 in multislice datasets and 6.7 ± 0.4 in ultrafast datasets (both reported as mean ± SD). Moreover, the Fourier maps obtained from both types of data (Figures 2D and 3D) show robust BOLD responses along the entire mouse visual pathway (V1, SC, LP and LGN) upon monocular flashing stimulation, mostly on the contralateral (right) side.

3.2. MP-PCA denoising of vendor reconstructed multislice fMRI images requires large sliding windows ($N \geq 250$) for good noise removal performance and improved functional activation mapping.

Figure 2B shows the images obtained after application of MP-PCA denoising with 5 different sliding windows on vendor reconstructed multislice images (Strategy A) of a representative scan. Upon denoising, noise levels were approximately at the same level as the noise observed in undenoised data (Figure 2A) when a small sliding window ([5 5 1], i.e., $N = 25$, where $N$ the total number of voxels within the sliding window) was used; noise levels were clearly reduced when larger sliding windows ([20 20 1], [40 40 1], [20 20 5] and [10 10 10], i.e., $N \geq 400$) were employed. Residual images (Figure 2C) do not contain edge effects or exhibit specific anatomical features and are thus shown to carry more of the estimated noised variance when larger windows were used. This effect becomes clearer when inspecting the plots portraying quantitative values of SNR gain and estimated noise variance in the residuals (Figure 2H,I) obtained upon denoising: in this particular scan, these were limited to 33% and 20%, respectively, when denoising was performed at the slice level (i.e. sliding window size = 1 in the slice dimension) and $N \leq 100$. The values increased to 64% and 35%, respectively, when using 3D windows (and $N \leq 125$), but only reached their maximal levels (>126% and >62%) when $N \geq 250$ (2D or 3D windows). Similar trends were observed in the BOLD Fourier maps: whereas the increase of FSA in activated voxels (Figure 2J) was limited to 6% for smaller windows (2D windows or $N \leq 125$), gains of up to 17% were reached with the larger windows (3D windows and $N \geq 250$), accompanied by a greater volume of activation (Figure 2E,K). As shown on the Fourier difference maps (Figure 2F), MP-PCA with larger windows highlighted activation in regions with BOLD effect, while leaving areas with only noise contribution approximately intact. Consistently with these results, single-voxel time-courses obtained when using a small denoising window ([5 5 1]) show almost identical profiles as the ones obtained from undenoised data (Figure 2G), whereas larger windows ([20 20 1] and [20 20 5]) clearly decrease the signal variation in the time-courses without affecting the functional changes obtained upon stimulation.

3.3. MP-PCA denoising of vendor reconstructed ultrafast fMRI images does not improve functional activation mapping even with the largest possible sliding window.

When Strategy A of MP-PCA denoising (i.e., directly denoising vendor-reconstructed data) was applied to ultrafast fMRI data, good noise removal performance was only achieved with the [60 60] and [74 96] windows, as shown in Figure 3. Specifically, in this particular scan, windows [2 2], [5 5], [20 20] and [40 40] ($N \leq 1600$) only provided a maximum SNR gain of 16% (Figure 3H), with image residuals only containing a maximum of 10% of the estimated noise variance (Figure 3I) and showing some structured appearance (Figure 3C). When a larger window of [60 60] ($N = 3600$) was employed, image residuals started to contain more of the estimated noise variance (51%) and did not show anatomical detail (Figure 3C), and an SNR gain of 81% could be achieved. These values further increased to 86% and 137%, respectively, when the largest possible denoising window ([74 96], corresponding to the entire image matrix) was used (Figure 3H,I). These improvements did not, however, translate into significant BOLD Fourier map enhancements. FSA values in activated regions remained in the same range as the values obtained from undenoised data (Figure 3E,F,J), with just a slightly more spatially extensive activation observed with the larger ($N \geq 3600$) windows (Figure 3E,K). Moreover, single-voxel time-courses did not show strikingly different profiles upon MP-PCA denoising when compared to the ones obtained from undenoised data (Figure 3G).

3.4. The performance of the MP-PCA denoising algorithm improves when ZF is removed from k-space.

Vendor k-space data can contain inherent ZF in typical EPI settings, which generate inter-voxel correlations in image space that entail additional signal components in the PCA domain. However, these zeros can be removed prior to denoising to avoid spatial correlations. Video S1 shows GE-EPI images from representative multislice and ultrafast fMRI scans after ZF removal from k-space and after application of MP-PCA denoising at the slice level with 8 different sliding windows (Strategy B). The movies reveal no apparent motion besides breathing or any type of artifact, neither before or after denoising. Moreover, fine features of the images observed in undenoised data are preserved after MP-PCA denoising.

When applying denoising Strategy B to multislice data, the resulting images already show highly enhanced SNR with the smallest possible window ([2 2]), in contrast to Strategy A, which is further increased with progressively larger windows (Video S1A and Figure 4A). Specifically, the average SNR gain increased from 64% to 134% for sliding windows between [2 2] and [5 5] ($N \leq 25$), respectively, and reached its maximum at 459% with a sliding window of [25 25] ($N = 625$), with relatively small deviations observed between individual scans (Figure 4D). As with Strategy A, residual maps do not contain recognizable anatomic features for all tested sliding windows (Figure 4B). Moreover, $\sigma$-normalized residuals are well approximated by a zero-centered normal distribution and have lower variance than unity, i.e., MP-PCA residuals have lower variance than the estimated noise variance, suggesting that the technique suppressed local signal fluctuations solely originating from thermal noise. Quantitatively, the variance of the residuals ranged from 45% (with the [2 2] window) to 96% (with the [25 25] window) of the estimated noise variance, with very small deviations registered between individual scans (Figure 4E). Moreover, it closely matched the expected variance reduction $\sigma_r^2$ obtained by truncating Gaussian noise components only (Figure S1B,C). Single-voxel time-courses exhibited in Figure 4C show attenuation of spurious signal fluctuations present in undenoised multislice data with denoising, with larger windows decreasing more the variations, without disturbing the signal changes observed during visual stimulation.

Similarly to what was observed in multislice data and contrary to what was observed with Strategy A, the application of denoising Strategy B on ultrafast data produced images with highly enhanced SNR already with the [2 2] ($N = 4$) window (Video S1B and Figure 5A). Particularly, MP-PCA denoising achieved an average SNR gain of 39% (Figure 5D) and image residuals already contained on average 33% of the estimated noise variance (Figure 5E) when using this window. However, unlike multislice data results, this effect faded-out with gradually larger windows and only came back with the [20 20] ($N = 400$) window (Video S1A, Figure 5A,D,E), with SNR gain values and percentage of estimated noise variance present in the residuals reaching a maximum of 73% and 48%, respectively, when the largest tested window ([25 25], i.e., $N$ = 625) was used. Consistently with these results, single-voxel time-courses obtained when using a [5 5] denoising window are very close to the ones obtained from undenoised data (Figure 5C), whereas smaller and larger windows ([2 2] and [20 20], respectively) slightly decreased the signal variation in the time-courses without affecting the functional changes obtained upon stimulation. Residuals show lack of anatomical structure and are well fitted by a zero-centered normal distribution (Figure 5B). As with multislice data, the variance of residuals (Figure S1I) followed the trend expected from MP-PCA theory (Figure S1H).

Regarding other MP-PCA denoising output metrics, the estimated noise variance $\sigma^2$ varied between individual scans and showed a convex trend with increasing window size in both types of data (Figure S1A,G), although much more accentuated in the ultrafast fMRI datasets. For a [2 2] sliding windows, the estimated $\sigma^2$ was higher in regions with smaller signal values (i.e., regions with large blood vessels) both in multislice (Figure S2B) and ultrafast (Figure S2E) data. For larger sliding windows ([5 5] and [20 20] in Figure S2B,E), $\sigma^2$ maps were less resolved and therefore smoother in both multislice and ultrafast data. When a [20 20] sliding window was used, $\sigma^2$ maps did not show any type of anatomical features but showed slightly lower values on the right and left extremes of the brain. Moreover, it can be noted that the sum of noise eigenvalues (Figure S1D,J), the number of retained "signal" components $P$ (Figure S1E,K) and the number of eliminated "noise" components $M' - P$ (Figure S1F,L) monotonically increase with larger window sizes (until the [20 20] sliding window) regardless of the data type (multislice or ultrafast) under study. For a [2 2] window, the number of signal components $P$ was higher in regions closer to the receiver coil (i.e., with higher signal values) in multislice data (Figure S2C), and practically homogeneous across the brain in ultrafast data (Figure S2F). As with $\sigma^2$, $P$ maps were less resolved and smoother in both types of data when using larger windows.

## 3.5. Larger denoising windows provide higher sensitivity to functional activation and increased spatial extent of activation.

Figure 6 shows the BOLD Fourier maps (and respective difference-to-undenoised maps) obtained from representative multislice and ultrafast fMRI datasets subjected to denoising Strategy B with 3 different sliding windows. Following the same trend as the previous results on the amount of noise removed and SNR gain achieved upon MP-PCA denoising, functional maps already show a clear increase in FSA values confined to regions of the visual pathway when a sliding window as small as [2 2] is employed (Figure 6A,F). Specifically, FSA values in activated regions increased on average 9% in multislice data and 7% in ultrafast data (Figure 6D,I) when using this window. In multislice data, these values further increased with progressively larger windows (Figure 6A,C), reaching a maximum average increase of 28% when a [20 20] window was used (Figure 6D). On the contrary, in ultrafast data, the FSA values in activated regions began to decrease with larger windows, reaching the undenoised data value range with a [5 5] window, and only started increasing again with the [20 20] window (Figure 6F,H), reaching a maximum average increase of 15% when a [25 25] window was used (Figure 6I).

Interestingly, the increase in sensitivity to functional activation provided by larger windows was accompanied by an increase in the number of activated voxels, as shown on the BOLD maps obtained after denoising with a [20 20] sliding window (Figure 6A,F). Specifically, the increase in the number of activated voxels reaches average values as high as 270% with a [10 10] sliding window in multislice data, and of 100% with a [25 25] sliding window in ultrafast data (Figure 6E,J). This increase in spatial extent of activation "spreads out" of the regions that were activated in the maps obtained from undenoised data (Figure 6C,H) and out of the ROIs of the visual pathway (Figure 6B,G). Moreover, it can be seen from the plots of Figure 6D,I that the increase of FSA in activated voxels starts to strongly vary between individual scans when larger windows ($N \geq 100$ in multislice data and $N \geq 400$ in ultrafast data) are employed. In some individual multislice scans, FSA values actually started to deviate from the average increasing pattern and to decrease when using windows equal or larger than [10 10] ($N \geq 100$).

After running a Pearson's correlation to test the relationship between the average FSA extracted from activated voxels of undenoised data BOLD maps (indicator of functional CNR) and the values of FSA increase for each sliding window (Figure S3A,B), we found a high correlation between these variables: on average $r(7) = 0.83$ with $p < 0.008$ for all windows in multislice fMRI data, and $r(7) = 0.78$, $p < 0.03$ for all windows in ultrafast fMRI data except the [10 10] and [15 15] ($p > 0.15$), where the FSA increase was the lowest (Figure 6I). On the contrary, we found a weak correlation between initial tSNR and FSA increase values for each sliding window, with an average $r(7) = -0.25$ ($p > 0.46$) in multislice data and $r(7) = -0.17$ ($p > 0.21$) in ultrafast data (Figure S3C,D).

### 3.6. Group results show higher activation to stimulation confined to regions of the visual pathway upon MP-PCA denoising with a [2 2] sliding window.

Individual data MP-PCA denoised with a sliding window of [2 2] and Strategy B were averaged to produce the group BOLD maps shown in Figure 7. Both in multislice (Figure 7A) and ultrafast (Figure 7C) group data results, MP-PCA highlighted activation in areas that were activated in the maps obtained from undenoised data, without substantially increasing the spatial extent of activation. Specifically, FSA values in activated voxels increased 15% in multislice data and 14% in ultrafast data in comparison to undenoised data results (Figure 6D,I). These increases were larger than the average increases of 9% and 7% obtained in individual BOLD maps, respectively. The increase in the volume of activation remained, however, approximately at the same level as the average increase observed in individual maps (Figure 6E,J). This remained true for gradually larger windows up to [5 5] in multislice data and [15 15] in ultrafast data. For larger windows, the increase in the number of activated voxels in the group maps became significantly greater than the average increase observed in individual maps.

### 3.7. Magnitude versus complex data denoising do not produce very different results.

The fact that MP-PCA denoising was applied to magnitude data in Strategy A and to complex data in Strategy B raises the question of whether the gains demonstrated with Strategy B are not only due to ZF removal from k-space prior to denoising but also partially related to the differences in the nature of the data that is being denoised. To address this question, we repeated Strategy A on the multislice and ultrafast fMRI acquisitions shown in Figures 2 and 3, respectively, i.e., MP-PCA denoising was applied immediately after outlier correction, but using complex data. Figures S4 and S5 display the results obtained after applying this modified Strategy A on multislice and ultrafast data, respectively. Despite some differences on the SNR gain values, noise removal performance and BOLD map metrics achieved with complex data denoising when compared to magnitude data denoising, both strategies produce very similar results in both types of data.

### 3.8. ZF can be re-introduced to k-space after MP-PCA denoising without significant changes on the results.

Given the fact that ZF removal from k-space undesirably reduces the initial "apparent" resolution of the acquired datasets, we repeated Strategy B in one multislice and one ultrafast fMRI acquisition but this time re-introducing the removed ZF regions to k-space after MP-PCA denoising, as shown on the pipeline of Figure S6A. It can be noted that the resulting Fourier maps (Figure S6B,D) follow the same trend as the maps shown in Figure 6A,F, i.e. an increase of FSA values in activated voxels accompanied by a greater volume of activation with increasing window size in multislice data (Figure S6C), and a U-shaped behavior of these variables (increase of FSA values and number of activated voxels) as a function of sliding window size in ultrafast data (Figure S6E).

3.9. Simulations show that MP-PCA denoising causes activation "spreading" and smoother functional maps.

As shown in Figure 8A, the Fourier map of GT simulated ultrafast fMRI data reveals a clear activation in the four different bilateral ROIs of the visual pathway (V1, SC, LP and LGN), with slightly higher FSA values found in the right hemisphere, i.e., where the BOLD response peaked later, in the [3.7,4.3] s interval after stimulus onset. When Gaussian white noise was added so that the average tSNR in the brain = 6.7 (Figure 8B), FSA values in activated regions markedly decreased, which led to the disappearance of LGN and LP from the thresholded map (although some spots with higher FSA are still visible on those regions in the map without threshold, especially in the right hemisphere). After applying MP-PCA denoising with a [2 2] sliding window on simulated noisy data (Figure 8C), the resulting BOLD map shows again a clear distinction between activated and non-activated regions and all the ROIs of the pathway pass the FSA threshold. Moreover, denoised images (shown in the background of the thresholded map) show a strong noise reduction and therefore a higher resemblance to the GT images than the noisy images, and error maps show lack of anatomy, indicating signal-preservation. However, when larger denoising windows are employed ([3 3] to [25 25]), although error maps show a smaller difference between denoised and GT images and FSA values in activated regions increase, functional maps become smoother and activation starts to "spread out" to neighboring voxels that were not activated in GT data. This effect is also visible in the single-voxel time-courses of Figure S7. Although an activated voxel in the right V1 (voxel 1 in Figure S7A) regained part of its initial functional CNR (Figure S7B) with denoising (Figure S7D), a voxel close to V1 and SC that was not activated in GT data (voxel 2 in Figure S7A) also started to exhibit an artifactual response to stimulation when a [10 10] sliding window was used (Figure S7E). A similar effect also occurred with another voxel even further away from activated regions (voxel 3 in Figure S7A), which also started to show an activation profile with the [20 20] denoising window (Figure S7F). After estimating the average temporal correlation of neighboring voxels through noise kernel calculation of brain image data (Figure S8), it was found that the functional point spread function (PSF) follows a Gaussian shape isotropically across space upon denoising (Figure S8B) and that its FWHM increases approximately linearly with denoising window side length (Figure S8C).

3.10. Activation "spreading" patterns are consistent if only the center voxel of each patch is kept during denoising.

To understand if the observed increase in spatial extent of activation could be partially related to the fact that data patches are being equally averaged after denoising, we repeated MP-PCA denoising on the simulated data (with tSNR = 6.7), but only keeping the center voxel of the patch at each slide. Results are displayed in Figure S9. Despite a slightly lower level of blur, activation "spreading" patterns and smoothing of functional maps upon larger denoising windows are still visible on the maps (Figure S9C).

3.11. Optimal MP-PCA sliding window for improved specificity of functional mapping depends on data tSNR and functional CNR.

While the [2 2] sliding window was shown to improve accuracy of functional measurements without generating false BOLD responses in non-activated regions in the simulated ultrafast datasets with tSNR = 6.7, different results were observed when different levels of noise were added to GT data, in particular, when tSNR = 3 (Figure S10) and when tSNR = 15 (Figure S11).

When a tSNR of 3 was imposed, the Fourier map became almost completely random and no voxel passed the threshold (Figure S10B). Although the right V1 and SC started to stand out from the rest of the brain in the BOLD map when a [2 2] sliding window was used (Figure S10C), the left V1 and SC only showed a clear activation when a [4 4] sliding window was employed. Moreover, the activation of LGN and LP is never spotted in the maps, except when activation has already spread to the entire upper part of the brain (with a [20 20] window). In contrast, when less noise was added to GT data (tSNR = 15), the BOLD map of noisy data still showed activation in almost the entire visual pathway (Figure S11B), and activation spreading to non-activated regions was already visible when a sliding window as small as [2 2] was used (Figure S11C).

The quantifications of these results are summarized in Figure 9. Particularly, it is shown that the increase of FSA in GT activated voxels (Figure 9A) and the increase in spatial extent of activation (Figure 9B) grow with larger windows at different rates, depending on the tSNR (the larger the tSNR, the larger the growth rate of those variables). Moreover, the Dice similarity coefficient (Figure 9C) calculated between each noisy or MP-PCA map and the GT BOLD map shows that the optimal sliding window for the most accurate spatial extent of activation depends on the tSNR: a maximum score of 0.51 was obtained for the [4 4] window when tSNR = 3, a maximum score of 0.76 for the [2 2] window when tSNR = 6.7, and a maximum score of 0.83 for undenoised data (i.e., MP-PCA denoising is not required at all) when tSNR = 15. Those optimal windows allow to yield a sensitivity of at least 0.70 in the maps (specifically, 0.70 for the [4 4] window when tSNR = 3, 0.91 for the [2 2] window when tSNR = 6.7, and 0.75 when data is not denoised when tSNR = 15)(Figure 9D), and to limit the FPR to values of 0.37 in the 1-pixel perimeter layer of the GT activation volume (Figure 9E), 0.10 in the 2-pixel perimeter layer (Figure 9F), and values lower than 0.05 in the 3- to 5-pixel perimeter layers (Figure 9G-I). Although larger denoising windows provide higher sensitivity, they also lead to higher FPR values.

Similar trends were observed for datasets that were simulated with smaller (Figure S12) or higher (Figure S13) values of percent signal changes of the BOLD signal. Even though the Fourier map of GT data was exactly the same for all conditions (Figures 8A, S12A and S13A), large differences began to appear once noise was added (Figures 8B, S12B and S13B). Moreover, although error maps and denoised images (Figures 8C, S12C and S13C) suggest that MP-PCA had the same performance in terms of removal of estimated noise, when smaller percent signal changes were imposed, activation spreading only became obvious when windows larger than [5 5] were employed (Figure S12C); when BOLD responses were higher, a [2 2] denoising window already revealed a notable amount of false positives (Figure S13C). Indeed, the plots with the Dice score obtained for each simulation and denoising window (Figure 10C) reveal that the optimal window size for improved mapping specificity varies with the amplitude of the BOLD response: a maximum score of 0.49 was obtained for the [3 3] window when BOLD amplitudes were distributed in the 0.10-0.70% range, and a maximum score of 0.93 was obtained for the undenoised data when responses were in the 2.75-4.25% range. Again, these were accompanied by reasonable levels of sensitivity (at least 0.51)(Figure 10D) and limited FPR (0.26 for $FPR_1$, 0.07 for $FPR_2$, and 0.02 for $FPR_3$, and 0 for $FPR_4$ and $FPR_5$)(Figure 10E-I). As with the results shown for simulations with different tSNR levels, different values of functional CNR resulted in different values of FSA (Figure 10A) and spatial extent (Figure 10B) increase for each of the denoising windows tested. Moreover, the larger the functional CNR, the larger the growth rate of those variables. When the functional CNR was the lowest (shown in purple in Figure 10A), it was actually possible to see a decreasing pattern of FSA increase values for windows larger than [10 10].

3.12. NORDIC PCA denoising also causes activation "spreading", but without producing smoother BOLD maps.

Figure 11 shows the Fourier maps obtained after applying NORDIC PCA denoising with 5 different windows to the simulated ultrafast fMRI dataset with a tSNR = 6.7 and 0.75-2.25% BOLD changes. When the phase-stabilization correction was not performed, it can be noted that activation started to "spread out" from the GT activated volume when windows larger than [5 5] were used (Figure 11C). The bilateral LGN and LP did not appear activated with any of the sliding windows tested, except when the entire brain was shown as activated (with a [40 40] window). Moreover, FSA values in GT activated voxels increased a maximum of 68% in comparison to undenoised data results when a [25 25] window was used (Figure 11E). Although the phase-stabilization correction approach increased the maximum FSA gain to 95% when a [10 10] window was used (Figure 11E), reduced the amount of spreading (Figure 11D,F), especially when using the [25 25] and [40 40] windows, and allowed to get activation in both LGN and LP with a [5 5] sliding window, it is still clearly visible that this denoising method generates BOLD responses outside of the original activation regions when large windows are employed. Nevertheless, it is noteworthy that NORDIC PCA denoising did not produce smoother maps as occurred with MP-PCA denoising (Figure 8C). Noise kernels estimated after NORDIC PCA denoising (Figure S8C) indicate a slight increase in spatiotemporal correlations when compared to undenoised data, although limited to a much smaller neighborhood than the ones obtained with MP-PCA denoising (Figure S8B). Moreover, on the contrary to MP-PCA, the functional PSFs are not isotropic across space, nor seem to follow a Gaussian shape in any direction.

3.13. Activation "spreading" patterns are consistent if a GLM analysis is used.

To test whether the FSA-based conclusions could be generalized to general fMRI studies, we performed a GLM analysis to the acquired multislice fMRI datasets and one of the ultrafast fMRI simulations. Figure S14A,C displays the GLM analysis results for the multislice fMRI acquisition shown in Figure 6A,C. Despite the higher detection of spurious negative responses using GLM analysis (Figure S14A) when compared to Fourier analysis results (Figure 6A), the increase in sensitivity to functional activation provided by larger windows was consistent between the methods (Figures 6C and S14C). Specifically, *t*-values in activated regions increased on average 5% in multislice data when using a [2 2] sliding window, reaching a maximum average increase of 14% when a [10 10] window was used (Figure S14D). Again, the increase of *t*-values in activated voxels started to strongly vary between individual scans when larger windows ($N \geq 100$) were employed (Figure S14D). In one of the scans, *t*-values actually started to decrease when a window as small as [2 2] was used. Moreover, the monotonic increase in the number of activated voxels provided by larger windows observed in BOLD Fourier maps (Figure 6A) was also visible in the BOLD GLM maps (Figure S14A). Specifically, the spatial extent of activation on these maps increased on average 127% when using a [10 10] window, reaching a maximum average 157% increase when the largest sliding window ([25 25]) was employed (Figure S14E). Therefore, despite growing at slightly lower rates when using a GLM analysis, the "spreading" patterns were consistent between both methods.

In agreement with these results, the GLM analysis results obtained for the simulated ultrafast data (Figure S14F-H) also show great similarities with the Fourier analysis results (Figure 8): slightly higher *t*-values found in the right hemisphere in the GT data map (Figure S14F), the disappearance of the left LGN and LP from the thresholded map when Gaussian white noise was added to the images (Figure S14G), the clearer distinction between activated and non-activated regions in BOLD maps after MP-PCA denoising with a [2 2] sliding window (Figure S14H), and, although at a slightly higher rate than Fourier analysis, the "spreading" pattern of activation with larger denoising windows (Figure S14H-J).

## 4. Discussion

MP-PCA denoising has been proven a successful method for the removal of thermal noise from redundant MRI data[28,30-33] and has allowed, for fMRI in particular, improved task-based and resting-state mapping[34-36]. However, nowadays, EPI is the most common acquisition method in fMRI experiments, providing data that is inherently weighted by gradient ramps which, together with the use of partial Fourier imaging techniques[39], introduce zeros (or otherwise a filtered reconstruction of missing data[53,54]) to k-space that generate spatial correlations that do not comply with the MP theory assumptions. Moreover, the projection onto the signal subspace upon MP-PCA denoising may lead to local "bleeding" of active signal components that result in activation "spreading". Here, we show that pre-clinical rodent fMRI mapping significantly benefits from MP-PCA denoising both when conventional multislice and ultrafast fMRI acquisitions are considered. Given MP-PCA denoising's efficacy in ultrafast fMRI data (which has an inherently lower SNR due to rapid acquisition and $T_1$ relaxation), our study suggests that higher temporal and spatial resolutions may be now more readily achievable. Image residuals did not show anatomical details and their histogram approximated the zero-centered normal distribution in all conditions tested (Figures 2C, 3C, 4B and 5B), suggesting that MP-PCA denoising conservatively suppressed only thermal (Gaussian) noise. Moreover, the variance of the residuals was systematically lower than the estimated noise variance (Figures 2I, 3I, 4E and 5E), which is in agreement with previous descriptions of the method[27,28]. Particularly, data variability due to noise can only be partially canceled by MP-PCA denoising, due to the noise corruption of the retained $P$ components.

Furthermore, we show that this technique has enhanced performance when the ZF regions of k-space are removed prior to denoising. Removing ZF from k-space (Strategy B) allowed to achieve a meaningful improvement in functional activation mapping in ultrafast fMRI data (Figure 6F,H) that was not possible whatsoever when denoising magnitude data obtained directly from the vendor's software (Strategy A)(Figure 3E,F), confirming that correlated noise can greatly decrease this technique's maximum capability. This was achieved at the cost of losing isotropic in-plane resolution, however, we have shown that the original ZF can be re-introduced to k-space after denoising and produce results in the same order of magnitude and with the same trend as the ones from Strategy B (Figure S6). Moreover, we have shown than the demonstrated gains are not due to the differences in the nature of the data that is being denoised in each strategy (Figures S4 and S5). Contrary to complex images whose noise is Gaussian-distributed, in magnitude image reconstructions the noise is either Rician or non-central $\chi^2$ distributed[55,56]. Therefore, at low SNR, the non-zero-mean thermal noise of magnitude data will become more apparent, reducing the efficacy of MP-PCA denoising. Given the possibility of complex-valued data extraction from pre-clinical scanners, we performed complex data denoising in Strategy B, as opposed to magnitude data denoising in Strategy A. However, we showed that denoising Strategy A was equally effective for magnitude and complex images, and therefore the nature of the data was not crucial here. Gudbjartsson and Patz[56] have shown that for SNR larger than two, the Rician noise distribution approximates to a Gaussian distribution. Here, both multislice and ultrafast fMRI data were characterized by a sufficiently large SNR to avoid the violation of the Gaussian assumption and performance degradation when denoising magnitude data. However, we emphasize that this result is particular to these datasets, therefore future studies should carry out an SNR analysis before performing MP-PCA denoising of magnitude data. It is noteworthy to mention that denoising of real-valued data is also a possibility, as this type of data is not superimposed by a noise floor but instead by a zero-mean Gaussian noise distribution, reducing the systematic bias introduced by the use of magnitude data[57,58]. We also note that including an outlier correction step before MP-PCA denoising might influence its performance. However, here only a very small percentage of datapoints was corrected in each scan. Specifically, we did not find any outlier in ultrafast data and, in multislice data, a maximum of 2 outliers/slice/scan (out of 340 time points) were found. Therefore, we do not expect major influences of outlier correction on the performance of denoising and subsequent analyses on our datasets.

Another important choice that should be taken into account when applying MP-PCA denoising is the sliding window size. Previous work[28] recommended using a window size similar or larger than the number of measurements, i.e., $N \gtrsim M$, for improved performance of MP-PCA denoising on dMRI data. However, in fMRI experiments, when high temporal resolutions are required, (1) a multislice type of acquisition may not always be possible, limiting the maximum denoising window size to the entire 2D matrix size, and (2) a higher number of measurements may need to be acquired to obtain sufficient temporal data. Under these conditions, as in our ultrafast fMRI data (where $N_{\max}$ = 7104 and $M$ = 9000), it might be impossible to meet this recommendation. Moreover, noise has to be spatially uniform within the sliding window (a critical MP-PCA assumption) for a precise estimation of the noise threshold. Here, we used a single-channel loop coil for signal reception, thus avoiding coil-driven spatial correlation of the noise[59]. However, even if we could assume that the noise level in our data is constant among all elements of the window and increase $N$ to be closer to $M$, it is also important to note that larger patches will become increasingly heterogeneous with respect to the different tissues with different HRFs that they cover. As such, they will deviate from being low-rank ($P$ will increase), making denoising less efficient. Diao et al.[36] recommended choosing a window within which BOLD fluctuations are likely to be correlated and thermal noise is not, this way avoiding the removal of genuine BOLD fluctuations and improving the sensitivity of the analysis to the latter. Here, we show that a substantial amount of noise was removed using Strategy A only when large sliding windows were employed (3D windows with $N \geq 250$ for multislice data, and 2D windows with $N \geq 3600$ for ultrafast data)(Figures 2H,I and 3H,I), while Strategy B was able to achieve similar results with much smaller windows (2D windows with $N \leq 25$)(Figures 4D,E and 5D,E). Although the best-performing windows of Strategy A seem to approach more the $N \gtrsim M$ recommendation than the ones of Strategy B, the tissue heterogeneity within each of those windows is remarkably high, especially when considering that 3D windows can cover tissues that are at a distance as high as (or even further than) 1 mm due to the voxel anisotropy. Hence, this could also explain in part the worse performance of MP-PCA denoising Strategy A in terms of functional activation mapping observed in ultrafast acquisitions. Interestingly, different behaviors were observed between multislice and ultrafast datasets in the SNR gain and estimated noise variance in the residuals with increasing window size in Strategy B. Whereas these metrics monotonically increased with larger window sizes in multislice data (Figure 4D,E), in ultrafast data they showed a U-shaped behavior, only starting to increase for windows with $N > 100$ (Figure 5D,E). The convex appearance of these plots is mostly due to the higher difference between the Casorati matrix dimensions, i.e., the higher value of $M$ used in ultrafast data (9000) in comparison to only 340 repetitions in multislice data for the same values of $N$, and to the number of "signal" components $P$ required for each type of data. Even if the sum of noise eigenvalues ($\sum_{i=P+1}^{M'} \lambda_i$), the number of retained "signal" components $P$ and the number of eliminated "noise" components $M' - P$ monotonically increased with larger window sizes (until the [20 20] sliding window) for both types of data acquired in this study (Figure S1D-F,J-L), the estimated noise variance $\sigma^2$, the expected variance reduction $\sigma_r^2$ and the expected SNR gain derived from the MP-PCA denoising theory also strongly depend on the ratios obtained from the sizes of the original and denoised Casorati matrices. In particular, the decreasing trend of $1/(M'-P)(N'-P)$ counterbalances the increasing trend of the sum of noise eigenvalues (Figure S1D,J) with larger window sizes and confers a convex shape to the estimated noise variance of both multislice and ultrafast data (Figure S1A,G). Whereas this is fully compensated by the monotonically increasing trend of $(M'-P)(N'-P)/(M'N')$ with larger window size in multislice data when estimating the expected variance reduction $\sigma_r^2$ (Figure S1B), in ultrafast data this ratio is actually convex, thereby giving an accentuated convex shape to the amount of variance reduced (Figure S1H) and consequently to the estimated SNR gain and the estimated noise variance in the residuals shown in Figure 5D,E. It should be noted that this U-shaped effect was also observed in the SNR gain and estimated noise variance in the residuals in the slice-wise denoised multislice (purple dots in Figure 2H,I) and ultrafast (Figure 3H,I) datasets in Strategy A. Although the $M$ and $N$ dimensions are the same when using this strategy, the number of retained $P$ signal components is significantly higher when compared to the values obtained with Strategy B due to the inter-voxel correlations generated by ZF. Therefore, the abovementioned ratios also differ between strategies, resulting in the different observed trends. When analysing the variation of the estimated $\sigma^2$ values across the brain (Figure S2B,E), we noted that the estimated $\sigma^2$ maps showed anatomical features when small sliding windows were employed in both types of data, probably reflecting some physiological noise interference. For larger windows, both $\sigma^2$ and $P$ maps (Figure S2C,F) became less resolved and smoother due to the larger coverage of these windows. Therefore, when a [20 20] sliding window was used, $\sigma^2$ maps did not show any type of anatomical features anymore, but showed slightly lower values on the right and left extremes of the brain, probably due to the stronger influence of lower SNR values from outside of the brain. Moreover, we noticed some ringing-like structured artifacts in the residual maps when ultrafast fMRI data was denoised with its under-performing windows, i.e., the ones that were not able to remove much of the estimated noise and therefore resulted in less SNR gain (e.g., window [5 5] using Strategy B in Figure 5B, and [2 2] or [20 20] using Strategy A in Figure 3C). A similar effect was also observed when denoising multislice data with a [5 5 1] window using Strategy A in Figure 2C. In all of these cases, there was a very small number of components being eliminated as "noise" (Figure S1L) when compared to the number of components required to be retained as "signal" (Figure S1K). Given the small number of components available to explain thermal noise, it is possible that in these cases the method is not able to properly estimate the right edge of the MP distribution and extract purely random noise, therefore eliminating some structure from the images. This structured appearance of the residuals was also visible when denoising with a very small [2 2] window (Figures 4B and 5B). In this case, even if the proportion of eliminated "noise" components was not as low, their absolute number was restricted to a maximum of 4. Therefore, the MP theorem might also have some difficulty to hold when very small windows are used.

We have shown that although larger windows ($N \geq 100$) allow a higher noise removal performance and higher FSA gains in some scans than small windows ($N \leq 25$) using Strategy B (Figure 6D,I), these are achieved with an increase in spatial extent of activation that spreads out of the regions activated in the maps obtained from undenoised data (Figure 6C,H). By performing simulations (Figures 8 and S7), we confirmed that this was in part associated with the denoising method itself and not only due to true voxel activation. Since MP-PCA denoising uses the $P$ signal components of the eigenspectrum to reconstruct all voxels present in the patch, it could be that components that carry fMRI activation patterns are being used to reconstruct not only active but also non-active voxels, causing this local bleeding of activation. In fact, all previous works that focused on the application of MP-PCA denoising to (resting-state and task-based) fMRI data[34-36] have shown a great value of the method for an increase in sensitivity to the BOLD fluctuations, however, the improvement in spatial specificity had never been explored in great detail. Moreover, the signal changes associated with activation are often too small to detect this "spreading" effect on commonly used residual or temporal autocorrelation analyses, suggesting that simulations that replicate the phenomenon in a controlled manner are important. In addition to the BOLD maps, we have shown in our simulations that the error maps relative to GT data also have a smooth appearance (Figure 8C), even if the residuals relative to undenoised data were apparently random (Figure 8C) and symmetric to the added noise (Figure 8B). As with the activation maps, we also attribute this result to the smoothing effect caused by denoising. Furthermore, we have shown that the observed "spreading" effect is not related to the fact that results of overlapping voxels from multiple patches are being averaged after denoising. Despite presenting an expected slightly lower level of blur, only keeping the center voxel of each patch during denoising also resulted in activation "spreading" patterns and smoother BOLD maps (Figure S9). To verify if this effect was restricted to MP-PCA or more general to other low-rank denoising methods, our simulations were extended to the more recent NORDIC PCA denoising[37,38], which has been reported to improve accuracy of functional maps without changes in spatial precision, global image smoothness or functional PSF. Although this method performs better with 3D isotropic windows with $N \approx 11 \cdot M$, which was impossible with our simulated data, we show that sliding windows as small as $N = 100$ already produce the same "spreading" effect observed with MP-PCA denoising (Figure 11). Nevertheless, NORDIC PCA denoising generated much narrower PSFs (Figure S8) and did not produce smoother functional maps in any condition as seen with MP-PCA denoising, making it the preferred method in this regard, and should therefore be further explored in the future.

Hitherto most of the studies in which MP-PCA or NORDIC PCA denoising were applied[28,34,38] were performed on human data, which is usually acquired with isotropic voxels in 3D, and used isotropic 3D patches for denoising. When resolution is isotropic, a 3D patch has a smaller spatial extent (i.e., a smaller maximum radius to the center of the patch) than a 2D patch with the same $N$, and can therefore achieve a smaller "spreading footprint" upon denoising. Whereas isotropic 3D patches can be highly adequate for these datasets, in anisotropic resolution data (such as our multislice data, where voxel resolution was 145x145x500 $\mu m^3$), a 3D isotropic window will be highly heterogeneous in the slice dimension with respect to the different HRFs that it covers and will thus induce a larger volume of artifactual activation. Anisotropic 3D patches are also a possibility, however, to be used in our multislice fMRI data, the slice dimension would have to be lowered to a point where the first two dimensions would already be close to the size of 2D patches with equal $N$. For this reason, we decided to just show the spreading effect for 2D patches in Strategy B; however, we note that activation spreading would also occur in the third window dimension if 3D patches had been used.

Simulations with three distinct levels of tSNR and BOLD response amplitudes showed that the optimal MP-PCA sliding window size for improved specificity of activation mapping (i.e., providing the highest Dice score) varies with data tSNR (Figure 9C) and functional CNR (Figure 10C), with lower tSNR/functional CNR data requiring larger windows (Figures S10 and S12) and higher tSNR/functional CNR data not even needing denoising (Figures S11 and S13). This could explain in part the variations in BOLD maps' metrics observed between each type of acquisition that was performed in our fMRI experiments (multislice data has higher tSNR than ultrafast data) and between their respective individual scans, which possibly have different percent signal changes of BOLD response due to intra- and inter-mouse variability in task sensitivity (Figure 6D,E,I,J). In fact, similarly to what was observed between individual scans in our fMRI experiments (Figure 6D,I), different values of functional CNR in the simulations resulted in different values of FSA increase for each of the denoising windows tested (Figure 10A). Moreover, when the functional CNR was the lowest (purple dots in Figure 10A), it was actually possible to see a decreasing pattern of FSA increase values for windows larger than [10 10], similar to the one reported for some individual scans of multislice fMRI data (Figure 6D). Given the strong indications for a relationship between functional CNR and FSA increase for each denoising window in our acquired data, we tested it through Pearson's correlation and indeed found a high correlation between these variables for both multislice and ultrafast data (Figure S3A,B). On the contrary, given the small variation of tSNR between individual scans of the same type (CV = 5% in multislice data and CV = 6% in ultrafast data), we did not expect tSNR to have an influence on the different FSA increase values obtained between individual scans. Nevertheless, we also tested this hypothesis and found a weak negative correlation between initial tSNR and FSA increase values for each sliding window (Figure S3C,D). Moreover, according to our simulations (Figure 9A), in case there would be a relationship between these variables, tSNR would have a positive correlation with FSA increase values, not negative. Although only variations of tSNR and functional CNR have been explored in the simulations, as well as the effect of ZF removal in real data, we do not exclude the possibility that other factors may also affect the efficiency of denoising, the rate of activation spreading and therefore the optimal MP-PCA sliding window size. Structured non-white noise (e.g., physiological noise) can have an impact on low-rank denoising performance, as these additional fluctuations will entail additional signal components in the PCA domain and make data less low-rank. One example of this effect is the fact that even though we set the same FOV, spatial and temporal resolution, repetitions, paradigm, tSNR and functional CNR in the simulations that we used in the acquired ultrafast fMRI datasets, there were different spreading behaviors between their results (Figure 6I,J versus Figures 9A,B and 10A,B), probably due to the presence of physiological noise in real data. Moreover, we note that the local bleeding effect reported in this work may not only involve "active" principal components but also any other retained $P$ signal components. Therefore, physiological noise can also be locally spread. However, the identification and quantification of such an effect would require other targeted analyses that are out of the scope of this work. We do not expect, however, an increase in the level of physiological noise upon denoising, as these methods are only focused on the reduction of thermal noise and therefore are only able at most to make other components easier to detect, not increase their absolute level.

Another aspect that warrants further discussion is the method chosen for BOLD mapping. Here, we chose the Fourier spectral analysis method since it takes advantage of the periodicity of the paradigm to detect active brain areas without having to assume or have a-priori knowledge of the HRFs[3,50]. Moreover, this method requires much less computational time than conventional GLM analyses, especially when analyzing data with a very large number of repetitions such as our ultrafast fMRI data. Additionally, GLM analyses are limited in the fact that they assume the same HRF for every voxel in the brain, which is far from being the truth. Although this does not pose a major problem for block design paradigms with long stimulation periods due to BOLD response saturation, when the blocks are very short in duration, as is the case of our ultrafast fMRI data (where the stimulus duration was only 1 s), the use of a single HRF shape can have a major impact in the detection of all activated areas. Despite the differences between these methods, it should be noted, however, that both of their estimated measures (FSA and *t*-values) depend on the noise level of the data: the FSA values due to the previous z-score normalization of the data, and the *t*-values due to the GLM standard error present in its equation. Under the MP-PCA assumption that absolute signal values are not affected by denoising, it was therefore expected that both of these measures would have a similar trend to SNR gain with larger window sizes. Indeed, after performing a GLM analysis to multislice fMRI data we found an increase in sensitivity to functional activation provided by larger windows (Figure S14D) similar to the one observed after Fourier analysis (Figure 6D). We note, however, that for some denoising windows and scans the FSA and *t*-values on activated voxels actually decreased, which we attribute to the smoothing effect caused by denoising on the absolute signal values. Although in theory the signal values should not be affected by denoising, by performing simulations we found that GT non-activated voxels can exhibit artifactual responses to stimulation with larger denoising windows (as shown in the single-voxel time-courses of Figure S7E,F). Just as the smoothing effect adds artifactual activity to non-activated voxels, it is also likely that it is reducing the activity in GT activated voxels, thus counteracting the increase in FSA/t-values generated by thermal noise reduction. Moreover, the artifactual responses generated in GT non-activated voxels keep the same shape and timings/frequencies as the neighboring GT activated voxels. Therefore, even though Fourier or GLM-based methods differ in the way they test if a voxel is active, both of them should be able to detect this type of spurious activation. For this reason, we did not anticipate major differences between these methods in the ability to detect activation spreading in the BOLD maps. As expected, the spatial extent of activation of BOLD maps produced by GLM analysis of multislice (Figure S14A,E) and simulated (Figure S14H) data increased with larger denoising windows. Therefore, we confirm the generalizability of our FSA-based conclusions to general fMRI studies employing conventional GLM analyses. We note that, due to component truncation, low-rank denoising methods alter the dimensionality of the data, resulting in fewer degrees of freedom for the voxel time-courses that are crucial in deriving statistical maps. Therefore, denoising can have an impact on the value and interpretation of statistical maps of functional activation. Given that the estimation of the effective degrees of freedom of an fMRI time-series is still a subject of on-going debate, not only in the field of denoising[60] but for fMRI in general (e.g., when using high temporal resolution sequences)[61,63], we did not attempt to address this issue here. Therefore, we avoided statistical *p*-value thresholds and used FSA thresholds instead (and *t*-value thresholds in case of GLM analysis) to simply provide a measure of activation relative to noise. Nevertheless, as a reference, we report the equivalent *p*-value (0.05) for undenoised data.

Finally, it should be noted that EPI's requirements of interpolation and gridding due to its non-blipped zig-zag trajectory introduce spatial correlations that reduce the independence of the noise in each voxel and are, therefore, a confounder in this study. Even if ZF is removed from k-space, the residual inter-voxel correlations still exist. Therefore, the application of MP-PCA denoising before gridding, i.e., in the rawest of data before any addition of spatial correlations, or the use of imaging sequences avoiding spatial interpolation should be considered in the future[64,65]. Moreover, data denoising in our work was simplified by the fact that we used a single-channel coil for data acquisition. However, nowadays, it is more common to use phased array coils that allow a good SNR over a larger field of view. Signal combination from multiple channels may as well alter the noise characteristics in which MP-PCA relies, thus it is important to guarantee that noise is uncorrelated and constant within the local neighborhood before denoising. The NORDIC approach offers a solution to this problem, by mapping the spatially varying noise to spatially identical noise through g-factor normalization[37], hence it should be explored in future studies. Lastly, data normalization to z-score using measurements of mean and standard deviation of the signal prior to Fourier spectrum calculation may not have been the ideal approach to compare MP-PCA denoised data results, as it assumes data distribution shape is identical between conditions. This may not be true for different denoising windows, as the amount of thermal (Gaussian) noise removed is different between each window and therefore some conditions might show a distribution that is less Gaussian. Although we do not anticipate large deviations from the observed trends of FSA values with increasing window size when using a different analysis methodology (there is always some retention of Gaussian noise in the data), future studies should consider other ways to compare MP-PCA denoising windows in terms of functional activation mapping without data distribution assumptions, such as a simple DC offset removal prior to fast Fourier transform computation and a nonparametric test to statistically compare FSA values at the paradigm's first harmonics with FSA values at higher frequencies of the spectrum.

# 5. Conclusions

In this paper, we developed an approach to use vendor data with MP-PCA to denoise pre-clinical task-based fMRI data series and improve sensitivity to BOLD activation in both conventional multislice "slow" as well as ultrafast fMRI. Removing the vendor's ZF from k-space data immediately after outlier correction and prior to complex data denoising decreases spatial correlations that limit this technique's maximum capability, thereby allowing the reduction of significant amounts of noise and improved functional activation mapping using sliding windows with little tissue heterogeneity. Larger sliding windows provide higher sensitivity to BOLD responses with both MP-PCA and NORDIC denoising, however, with significant activation "spreading" and increases in FPR due to the local bleeding of active signal components. The optimal sliding window for each experiment will depend on data tSNR and functional CNR, which should be considered before denoising. Our results bode well for dramatically enhancing spatial and/or temporal resolution in future fMRI work, while taking into account these sensitivity/specificity trade-offs of low-rank denoising methods.

## Code availability

MP-PCA denoising code is available on GitHub at https://github.com/sunenj/MP-PCA-Denoising.

## Declaration of Competing Interests

None

## Credit authorship contribution

**Francisca F. Fernandes**: Conceptualization, Methodology, Formal analysis, Investigation, Data Curation, Writing – original draft, Writing – review & editing, Visualization. **Jonas L. Olesen**: Methodology, Software, Writing – review & editing. **Sune N. Jespersen**: Methodology, Software, Writing – review & editing, Supervision, Funding acquisition. **Noam Shemesh:** Conceptualization, Methodology, Resources, Writing – original draft, Writing – review & editing, Supervision, Project administration, Funding acquisition.

## Acknowledgements

The authors thank Rita Gil for support with ultrafast data acquisition and Dr. Cristina Chavarrías for advice on simulations. This study was funded in part by the European Research Council (ERC) (agreement No. 679058) and by Champalimaud Foundation. The authors acknowledge the vivarium of the Champalimaud Centre for the Unknow, a facility of CONGENTO which is a research infrastructure co-financed by Lisboa Regional Operational Programme (Lisboa2020), under the PORTUGAL 2020 Partnership Agreement, through the European Regional Development Fund (ERDF) and Fundação para a Ciência e Tecnologia (Portugal) under the project LISBOA-01-0145-FEDER-022170.

# References

[1] X. Yu, C. Qian, D. Chen, S. J. Dodd, and A. P. Koretsky, "Deciphering laminar-specific neural inputs with line-scanning fMRI," *Nat Methods*, vol. 11, no. 1, pp. 55–58, Jan. 2014, doi: 10.1038/nmeth.2730.

[2] H. L. Lee, Z. Li, E. J. Coulson, and K. H. Chuang, "Ultrafast fMRI of the rodent brain using simultaneous multi-slice EPI," *Neuroimage*, vol. 195, pp. 48–58, Jul. 2019, doi: 10.1016/j.neuroimage.2019.03.045.

[3] R. Gil, F. F. Fernandes, and N. Shemesh, "Neuroplasticity-driven timing modulations revealed by ultrafast functional magnetic resonance imaging," *Neuroimage*, vol. 225, p. 117446, Jan. 2021, doi: 10.1016/j.neuroimage.2020.117446.

[4] W. B. Jung, G. H. Im, H. Jiang, and S.-G. Kim, "Early fMRI responses to somatosensory and optogenetic stimulation reflect neural information flow," *Proc Natl Acad Sci U S A*, vol. 118, no. 11, p. e2023265118, Mar. 2021, doi: 10.1073/pnas.2023265118.

[5] D. H. Lim, J. LeDue, M. H. Mohajerani, M. P. Vanni, and T. H. Murphy, "Optogenetic approaches for functional mouse brain mapping," *Front Neurosci*, vol. 7, p. 54, Apr. 2013, doi: 10.3389/fnis.2013.00054.

[6] L. Tian *et al.*, "Imaging neural activity in worms, flies and mice with improved GCaMP calcium indicators," *Nat Methods*, vol. 6, no. 12, pp. 875–881, Dec. 2009, doi: 10.1038/nmeth.1398.

[7] H. Dana *et al.*, "High-performance calcium sensors for imaging activity in neuronal populations and microcompartments," *Nat Methods*, vol. 16, no. 7, pp. 649–657, Jun. 2019, doi: 10.1038/s41592-019-0435-6.

[8] E. Jonckers, D. Shah, J. Hamaide, M. Verhoye, and A. van der Linden, "The power of using functional fMRI on small rodents to study brain pharmacology and disease," *Front Pharmacol*, vol. 6, no. OCT, 2015, doi: 10.3389/FPHAR.2015.00231.

[9] D. Ratering, C. Baltes, J. Nordmeyer-Massner, D. Marek, and M. Rudin, "Performance of a 200-MHz cryogenic RF probe designed for MRI and MRS of the murine brain," *Magn Reson Med*, vol. 59, no. 6, pp. 1440–1447, Jun. 2008, doi: 10.1002/mrm.21629.

[10] D. Nunes, R. Gil, and N. Shemesh, "A rapid-onset diffusion functional MRI signal reflects neuromorphological coupling dynamics," *Neuroimage*, vol. 231, p. 117862, May 2021, doi: 10.1016/j.neuroimage.2021.117862.

[11]  L. I. Rudin, S. Osher, and E. Fatemi, "Nonlinear total variation based noise removal algorithms," *Physica D*, vol. 60, no. 1–4, pp. 259–268, Nov. 1992, doi: 10.1016/0167-2789(92)90242-F.

[12]  F. Knoll, K. Bredies, T. Pock, and R. Stollberger, "Second order total generalized variation (TGV) for MRI," *Magn Reson Med*, vol. 65, no. 2, pp. 480–491, Feb. 2011, doi: 10.1002/mrm.22595.

[13]  P. Coupe, P. Yger, S. Prima, P. Hellier, C. Kervrann, and C. Barillot, "An optimized blockwise nonlocal means denoising filter for 3-D magnetic resonance images," *IEEE Trans Med Imaging*, vol. 27, no. 4, pp. 425–441, Apr. 2008, doi: 10.1109/tmi.2007.906087.

[14]  J. v. Manjón, J. Carbonell-Caballero, J. J. Lull, G. García-Martí, L. Martí-Bonmatí, and M. Robles, "MRI denoising using non-local means," *Med Image Anal*, vol. 12, no. 4, pp. 514–523, Aug. 2008, doi: 10.1016/J.MEDIA.2008.02.004.

[15]  J. v. Manjón, P. Coupé, L. Martí-Bonmatí, D. L. Collins, and M. Robles, "Adaptive non-local means denoising of MR images with spatially varying noise levels," *Journal of Magnetic Resonance Imaging*, vol. 31, no. 1, pp. 192–203, Jan. 2010, doi: 10.1002/jmri.22003.

[16]  S. G. Kafali, T. Çukur, and E. U. Saritas, "Phase-correcting non-local means filtering for diffusion-weighted imaging of the spinal cord," *Magn Reson Med*, vol. 80, no. 3, pp. 1020–1035, Sep. 2018, doi: 10.1002/mrm.27105.

[17]  S. M. Smith and J. M. Brady, "SUSAN - A new approach to low level image processing," *Int J Comput Vis*, vol. 23, no. 1, pp. 45–78, 1997, doi: 10.1023/A:1007963824710.

[18]  C. Triantafyllou, R. D. Hoge, and L. L. Wald, "Effect of spatial smoothing on physiological noise in high-resolution fMRI," *Neuroimage*, vol. 32, no. 2, pp. 551–557, Aug. 2006, doi: 10.1016/j.neuroimage.2006.04.182.

[19]  E. K. Molloy, M. E. Meyerand, and R. M. Birn, "The influence of spatial resolution and smoothing on the detectability of resting-state and task fMRI," *Neuroimage*, vol. 86, pp. 221–230, Feb. 2014, doi: 10.1016/j.neuroimage.2013.09.001.

[20]  K. N. Kay, A. Rokem, J. Winawer, R. F. Dougherty, and B. A. Wandell, "GLMdenoise: a fast, automated technique for denoising task-based fMRI data," *Front Neurosci*, vol. 7, p. 247, Dec. 2013, doi: 10.3389/fnins.2013.00247.

[21] V. M. Pai, S. Rapacchi, P. Kellman, P. Croisille, and H. Wen, "PCATMIP: enhancing signal intensity in diffusion-weighted magnetic resonance imaging," *Magn Reson Med*, vol. 65, no. 6, pp. 1611–1619, Jun. 2011, doi: 10.1002/mrm.22748.

[22] J. v. Manjón, P. Coupé, L. Concha, A. Buades, D. L. Collins, and M. Robles, "Diffusion Weighted Image Denoising Using Overcomplete Local PCA," *PLoS One*, vol. 8, no. 9, p. e73021, Sep. 2013, doi: 10.1371/journal.pone.0073021.

[23] J. v. Manjón, P. Coupé, and A. Buades, "MRI noise estimation and denoising using non-local PCA," *Med Image Anal*, vol. 22, no. 1, pp. 35–47, May 2015, doi: 10.1016/j.media.2015.01.004.

[24] C. M. Sonderer and N.-K. Chen, "Improving the Accuracy, Quality, and Signal-To-Noise Ratio of MRI Parametric Mapping Using Rician Bias Correction and Parametric-Contrast-Matched Principal Component Analysis (PCM-PCA)," *Yale J Biol Med*, vol. 91, no. 3, p. 214, Sep. 2018, Accessed: Jan. 20, 2022. [Online]. Available: /pmc/articles/PMC6153621/

[25] Y. Behzadi, K. Restom, J. Liau, and T. T. Liu, "A component based noise correction method (CompCor) for BOLD and perfusion based fMRI," *Neuroimage*, vol. 37, no. 1, pp. 90–101, Aug. 2007, doi: 10.1016/j.neuroimage.2007.04.042.

[26] C. Caballero-Gaudes and R. C. Reynolds, "Methods for cleaning the BOLD fMRI signal," *Neuroimage*, vol. 154, pp. 128–149, Jul. 2017, doi: 10.1016/j.neuroimage.2016.12.018.

[27] J. Veraart, E. Fieremans, and D. S. Novikov, "Diffusion MRI noise mapping using random matrix theory," *Magn Reson Med*, vol. 76, no. 5, pp. 1582–1593, Nov. 2016, doi: 10.1002/mrm.26059.

[28] J. Veraart, D. S. Novikov, D. Christiaens, B. Ades-aron, J. Sijbers, and E. Fieremans, "Denoising of diffusion MRI using random matrix theory," *Neuroimage*, vol. 142, pp. 394–406, Nov. 2016, doi: 10.1016/j.neuroimage.2016.08.016.

[29] V. A. Marčenko and L. A. Pastur, "Distribution of eigenvalues for some sets of random matrices," *Mathematics of the USSR-Sbornik*, vol. 1, no. 4, pp. 457–483, Apr. 1967, doi: 10.1070/SM1967V001N04ABEH001994.

[30] I. M. Adanyeguh *et al.*, "Autosomal dominant cerebellar ataxias: Imaging biomarkers with high effect sizes," *Neuroimage Clin*, vol. 19, pp. 858–867, Jun. 2018, doi: 10.1016/j.nicl.2018.06.011.

[31]  E. T. McKinnon, J. A. Helpern, and J. H. Jensen, "Modeling white matter microstructure with fiber ball imaging," *Neuroimage*, vol. 176, pp. 11–21, Aug. 2018, doi: 10.1016/j.neuroimage.2018.04.025.

[32]  F. Grussu *et al.*, "Multi-parametric quantitative in vivo spinal cord MRI with unified signal readout and image denoising," *Neuroimage*, vol. 217, p. 116884, Aug. 2020, doi: 10.1016/j.neuroimage.2020.116884.

[33]  M. D. Does *et al.*, "Evaluation of principal component analysis image denoising on multi-exponential MRI relaxometry," *Magn Reson Med*, vol. 81, no. 6, pp. 3503–3514, Jun. 2019, doi: 10.1002/mrm.27658.

[34]  B. Ades-Aron *et al.*, "Improved Task-based Functional MRI Language Mapping in Patients with Brain Tumors through Marchenko-Pastur Principal Component Analysis Denoising," *Radiology*, vol. 298, no. 2, pp. 365–373, Feb. 2021, doi: 10.1148/radiol.2020200822.

[35]  B. M. Adhikari *et al.*, "A resting state fMRI analysis pipeline for pooling inference across diverse cohorts: an ENIGMA rs-fMRI protocol," *Brain Imaging Behav*, vol. 13, no. 5, pp. 1453–1467, Oct. 2019, doi: 10.1007/s11682-018-9941-x.

[36]  Y. Diao, T. Yin, R. Gruetter, and I. O. Jelescu, "PIRACY: An Optimized Pipeline for Functional Connectivity Analysis in the Rat Brain," *Front Neurosci*, vol. 15, p. 602170, Mar. 2021, doi: 10.3389/fnis.2021.602170.

[37]  S. Moeller *et al.*, "NOise reduction with DIstribution Corrected (NORDIC) PCA in dMRI with complex-valued parameter-free locally low-rank processing," *Neuroimage*, vol. 226, p. 117539, Feb. 2021, doi: 10.1016/j.neuroimage.2020.117539.

[38]  L. Vizioli *et al.*, "Lowering the thermal noise barrier in functional brain mapping with magnetic resonance imaging," *Nat Commun*, vol. 12, p. 5181, Aug. 2021, doi: 10.1038/s41467-021-25431-8.

[39]  J. J. M. Cuppen, J. P. Groen, and J. Konijn, "Magnetic resonance fast Fourier imaging," *Med Phys*, vol. 13, no. 2, pp. 248–253, Mar. 1986, doi: 10.1118/1.595905.

[40]  M. S. Grubb and I. D. Thompson, "Quantitative Characterization of Visual Response Properties in the Mouse Dorsal Lateral Geniculate Nucleus," *J Neurophysiol*, vol. 90, no. 6, pp. 3594–3607, Dec. 2003, doi: 10.1152/jn.00699.2003.

[41]  C. M. Niell and M. P. Stryker, "Highly Selective Receptive Fields in Mouse Visual Cortex," *Journal of Neuroscience*, vol. 28, no. 30, pp. 7520–7536, Jul. 2008, doi: 10.1523/jneurosci.0623-08.2008.

[42] L. Wang, R. Sarnaik, K. Rangarajan, X. Liu, and J. Cang, "Visual Receptive Field Properties of Neurons in the Superficial Superior Colliculus of the Mouse," *Journal of Neuroscience*, vol. 30, no. 49, pp. 16573–16584, Dec. 2010, doi: 10.1523/jneurosci.3305-10.2010.

[43] A. Niranjan, I. N. Christie, S. G. Solomon, J. A. Wells, and M. F. Lythgoe, "fMRI mapping of the visual system in the mouse brain with interleaved snapshot GE-EPI," *Neuroimage*, vol. 139, pp. 337–345, Oct. 2016, doi: 10.1016/j.neuroimage.2016.06.015.

[44] A. Niranjan, B. Siow, M. Lythgoe, and J. Wells, "High temporal resolution BOLD responses to visual stimuli measured in the mouse superior colliculus," *Matters (Zur)*, vol. 3, no. 2, Feb. 2017, doi: 10.19185/matters.201701000001.

[45] T. N. A. Dinh, W. B. Jung, H. J. Shim, and S. G. Kim, "Characteristics of fMRI responses to visual stimulation in anesthetized vs. awake mice," *Neuroimage*, vol. 226, p. 117542, Feb. 2021, doi: 10.1016/j.neuroimage.2020.117542.

[46] J. M. Adamczak, T. D. Farr, J. U. Seehafer, D. Kalthoff, and M. Hoehn, "High field BOLD response to forepaw stimulation in the mouse," *Neuroimage*, vol. 51, no. 2, pp. 704–712, Jun. 2010, doi: 10.1016/j.neuroimage.2010.02.083.

[47] B. Pradier *et al.*, "Combined resting state-fMRI and calcium recordings show stable brain states for task-induced fMRI in mice under combined ISO/MED anesthesia," *Neuroimage*, vol. 245, p. 118626, Dec. 2021, doi: 10.1016/j.neuroimage.2021.118626.

[48] J. L. Olesen, A. Ianus, L. Østergaard, N. Shemesh, and S. N. Jespersen, "Tensor denoising of multidimensional MRI data," *Magn Reson Med*, pp. 1–13, Oct. 2022, doi: 10.1002/mrm.29478.

[49] E. S. Lein *et al.*, "Genome-wide atlas of gene expression in the adult mouse brain," *Nature*, vol. 445, no. 7124, pp. 168–176, Jan. 2007, doi: 10.1038/nature05453.

[50] D. Nunes, A. Ianus, and N. Shemesh, "Layer-specific connectivity revealed by diffusion-weighted functional MRI in the rat thalamocortical pathway," *Neuroimage*, vol. 184, pp. 646–657, Jan. 2019, doi: 10.1016/j.neuroimage.2018.09.050.

[51] L. R. Huber *et al.*, "LayNii: A software suite for layer-fMRI," *Neuroimage*, vol. 237, p. 118091, Aug. 2021, doi: 10.1016/j.neuroimage.2021.118091.

[52] Z. Fang, N. van Le, M. K. Choy, and J. H. Lee, "High spatial resolution compressed sensing (HSPARSE) functional MRI," *Magn Reson Med*, vol. 76, no. 2, pp. 440–455, Aug. 2016, doi: 10.1002/mrm.25854.

[53] E. M. Haacke, E. D. Lindskogj, and W. Lin, "A fast, iterative, partial-fourier technique capable of local phase recovery," *Journal of Magnetic Resonance (1969)*, vol. 92, no. 1, pp. 126–145, Mar. 1991, doi: 10.1016/0022-2364(91)90253-P.

[54] Y. Xu and E. M. Haacke, "Partial Fourier imaging in multi-dimensions: A means to save a full factor of two in time," *Journal of Magnetic Resonance Imaging*, vol. 14, no. 5, pp. 628–635, Nov. 2001, doi: 10.1002/jmri.1228.

[55] R. M. Henkelman, "Measurement of signal intensities in the presence of noise in MR images," *Med Phys*, vol. 12, no. 2, pp. 232–233, 1985, doi: 10.1118/1.595711.

[56] H. Gudbjartsson and S. Patz, "The Rician distribution of noisy MRI data," *Magn Reson Med*, vol. 34, no. 6, pp. 910–914, 1995, doi: 10.1002/mrm.1910340618.

[57] C. Eichner *et al.*, "Real diffusion-weighted MRI enabling true signal averaging and increased diffusion contrast," *Neuroimage*, vol. 122, pp. 373–384, Nov. 2015, doi: 10.1016/j.neuroimage.2015.07.074.

[58] Q. Fan *et al.*, "Axon diameter index estimation independent of fiber orientation distribution using high-gradient diffusion MRI," *Neuroimage*, vol. 222, p. 117197, Nov. 2020, doi: 10.1016/j.neuroimage.2020.117197.

[59] O. Dietrich, J. G. Raya, S. B. Reeder, M. Ingrisch, M. F. Reiser, and S. O. Schoenberg, "Influence of multichannel combination, parallel imaging and other reconstruction techniques on MRI noise characteristics," *Magn Reson Imaging*, vol. 26, no. 6, pp. 754–762, Jul. 2008, doi: 10.1016/j.mri.2008.02.001.

[60] A. X. Patel and E. T. Bullmore, "A wavelet-based estimator of the degrees of freedom in denoised fMRI time series for probabilistic testing of functional connectivity and brain graphs," *Neuroimage*, vol. 142, pp. 14–26, Nov. 2016, doi: 10.1016/j.neuroimage.2015.04.052.

[61] F. Kruggel, M. Pélégrini-Issac, and H. Benali, "Estimating the effective degrees of freedom in univariate multiple regression analysis," *Med Image Anal*, vol. 6, no. 1, pp. 63–75, Mar. 2002, doi: 10.1016/s1361-8415(01)00052-4.

[62] M. M. Monti, "Statistical Analysis of fMRI Time-Series: A Critical Review of the GLM Approach," *Front Hum Neurosci*, vol. 5, p. 28, Mar. 2011, doi: 10.3389/fnhum.2011.00028.

[63] J. E. Chen, J. R. Polimeni, S. Bollmann, and G. H. Glover, "On the analysis of rapidly sampled fMRI data," *Neuroimage*, vol. 188, pp. 807–820, Mar. 2019, doi: 10.1016/j.neuroimage.2019.02.008.

[64] G. Lemberskiy, S. Baete, J. Veraart, T. M. Shepherd, E. Fieremans, and D. S. Novikov, "Achieving sub-mm clinical diffusion MRI resolution by removing noise during reconstruction using random matrix theory," in *Proc. Intl. Soc. Mag. Reson. Med. 27*, 2019, p. 0770.

[65] G. Lemberskiy, S. Baete, J. Veraart, T. M. Shepherd, E. Fieremans, and D. S. Novikov, "MRI below the noise floor," in *Proc. Intl. Soc. Mag. Reson. Med. 28*, 2020, p. 3451.

# Figures

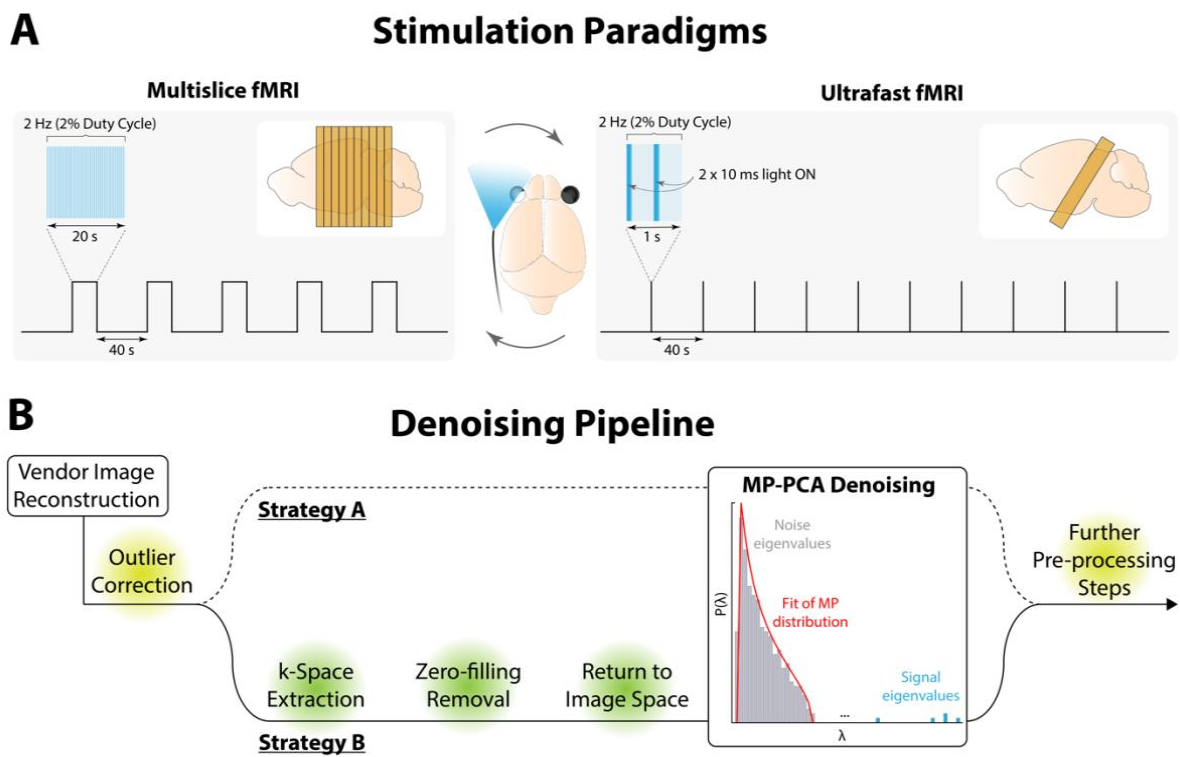

**Figure 1 – Stimulation paradigms and denoising flowchart. (A)** Stimulation paradigms used for multislice (*left*) and ultrafast (*right*) fMRI acquisitions, and respective slice positioning. Flashing blue light stimulation at 2 Hz frequency and 10 ms pulse width was applied to the left eye while the right eye was covered (*middle*). **(B)** Denoising strategies used during pre-processing. Vendor image data were either MP-PCA denoised immediately after outlier correction (Strategy A, dashed line), or transformed into k-space for ZF removal before denoising (Strategy B, solid line).

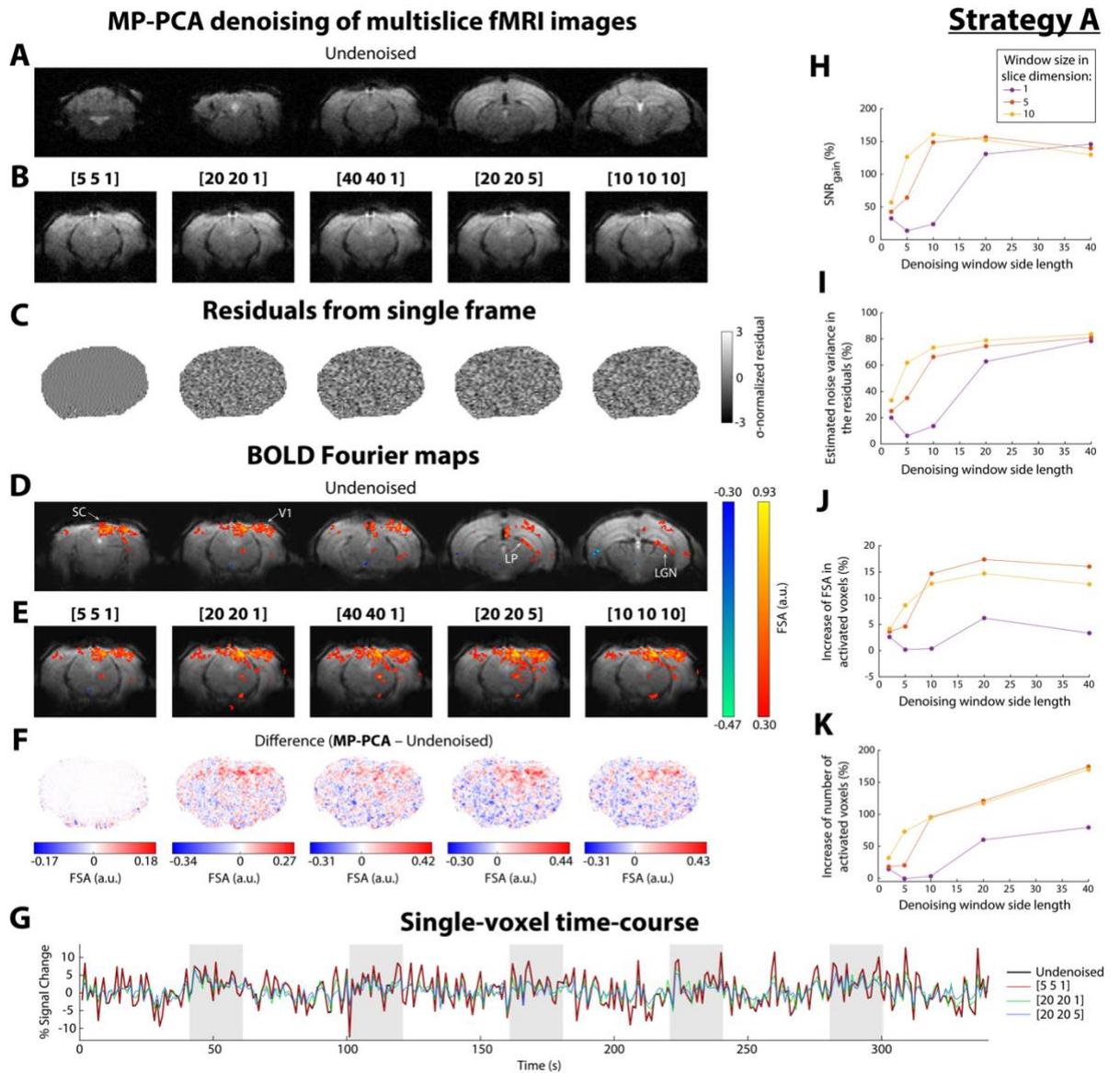

**Figure 2 – MP-PCA denoising of multislice fMRI data reconstructed by vendor software (Strategy A). (A)** Single GE-EPI image obtained from a representative multislice fMRI acquisition, before (5 out of 10 slices shown) and **(B)** after (1 out of 10 slices shown) MP-PCA denoising with 5 different sliding windows ([5 5 1], [20 20 1], [40 40 1], [20 20 5] and [10 10 10]) immediately after outlier correction. **(C)** Map of $\sigma$-normalized residuals of a single frame obtained after MP-PCA denoising with these 5 windows. **(D)** BOLD Fourier maps obtained before and **(E)** after MP-PCA denoising with these 5 windows. Maps are thresholded with a minimum FSA at paradigm's fundamental frequency = 0.3 and a minimum cluster size = 10. **(F)** Difference between the functional maps shown in (E) and (D). **(G)** Single-voxel detrended time-courses before (black line) and after MP-PCA denoising with 3 different sliding windows ([5 5 1] in red, [20 20 1] in green and [20 20 5] in blue). Grey areas represent the periods of visual stimulation. **(H)** Average brain SNR gain, **(I)** percentage of estimated noise variance explained by the residuals, **(J)** percentage increase of FSA in activated voxels, and **(K)** increase in spatial extent of activation, obtained after MP-PCA denoising with 15 different sliding windows. In particular, sliding window size varied between [2 2], [5 5], [10 10], [20 20] and [40 40] in the row and columns dimensions, and between 1 (in purple), 5 (in orange) and 10 (in yellow) in the slice dimension.

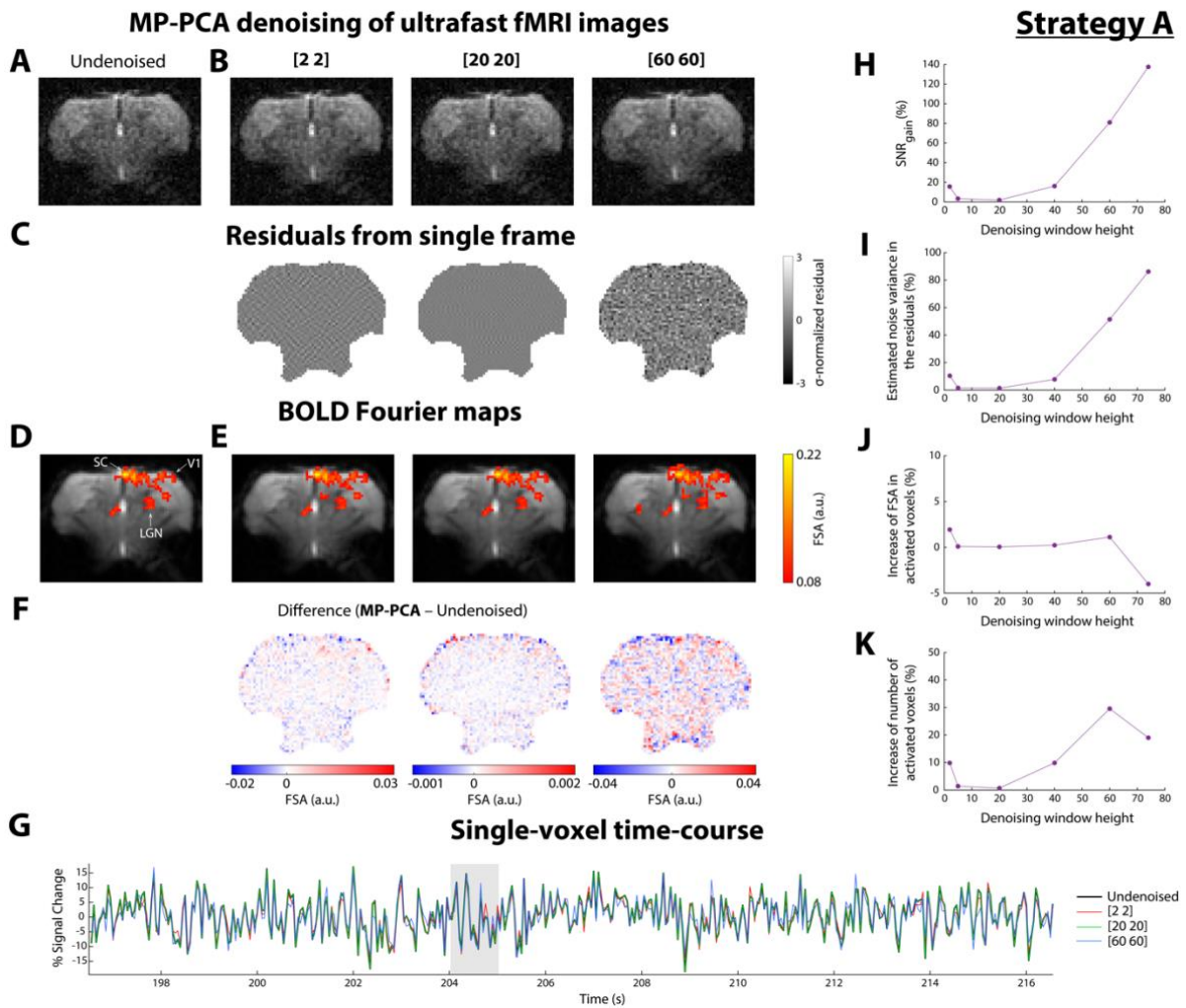

**Figure 3 – MP-PCA denoising of ultrafast fMRI data reconstructed by vendor software (Strategy A). (A)** GE-EPI image obtained from a representative ultrafast fMRI acquisition, before and **(B)** after MP-PCA denoising with 3 different sliding windows ([2 2], [20 20] and [60 60]) immediately after outlier correction. **(C)** Map of $\sigma$-normalized residuals of a single frame obtained after MP-PCA denoising with these 3 windows. **(D)** BOLD Fourier maps obtained before and **(E)** after MP-PCA denoising with these 3 windows. Maps are thresholded with a minimum sum of FSA at paradigm's fundamental frequency and two following harmonics = 0.08 and a minimum cluster size = 8. **(F)** Difference between the functional maps shown in (E) and (D). **(G)** Single-voxel detrended time-courses (only repetitions 3931 to 4331 are shown) before (black line) and after MP-PCA denoising with 3 different sliding windows ([2 2] in red, [20 20] in green and [60 60] in blue). The grey area represents the fifth period of visual stimulation. **(H)** Average brain SNR gain, **(I)** percentage of estimated noise variance explained by the residuals, **(J)** percentage increase of FSA in activated voxels, and **(K)** increase in spatial extent of activation, obtained after MP-PCA denoising with 6 different sliding windows.

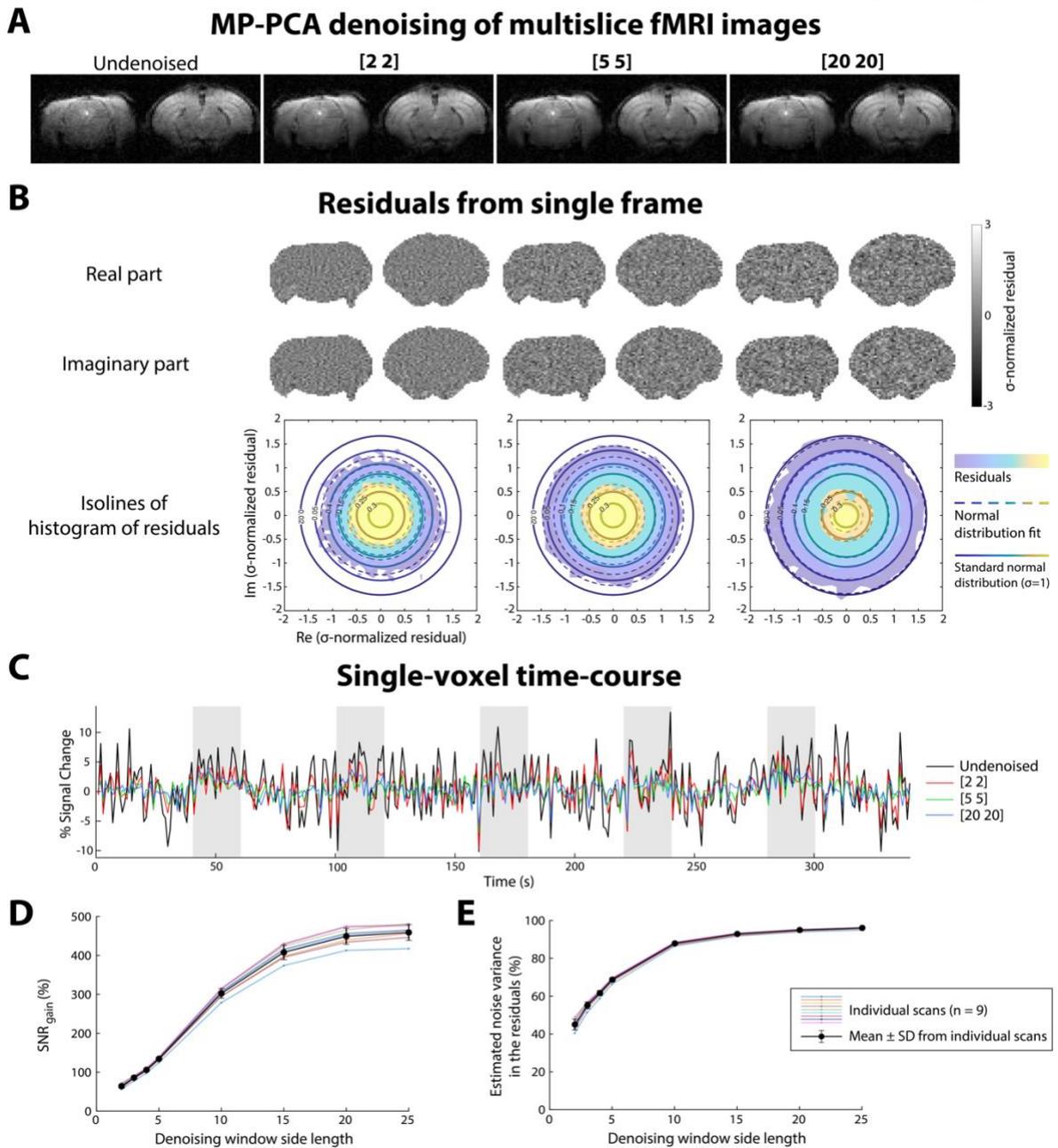

**Figure 4 – Residuals, single-voxel time-courses and SNR gain of multislice fMRI data after MP-PCA denoising (Strategy B). (A)** GE-EPI image of a representative multislice fMRI acquisition (2 out of 10 slices shown), before and after MP-PCA denoising at the slice level with 3 different sliding windows ([2 2], [5 5] and [20 20]). ZF was removed from k-space prior to data denoising. **(B)** Maps of real (*top*) and imaginary (*middle*) part of $\sigma$-normalized residuals of a single frame obtained after MP-PCA denoising with these 3 windows, together with filled contour plots containing isolines of the histogram of those residuals (*bottom*). Dashed lines represent the approximated normal distribution for the residuals whereas solid lines represent the standard normal distribution (unitary variance). **(C)** Single-voxel detrended time-courses before (black line) and after MP-PCA denoising with 3 different sliding windows ([2 2] in red, [5 5] in green and [20 20] in blue). Grey areas represent the periods of visual stimulation. **(D)** Average brain SNR gain across 8 different sliding windows and n = 9 different scans (represented by each color plot) of multislice fMRI data. The black dots with error bars

and black lines represent the mean ± SD results obtained from the individual scans. **(E)** Percentage of estimated noise variance explained by the residuals across the same sliding windows and scans.

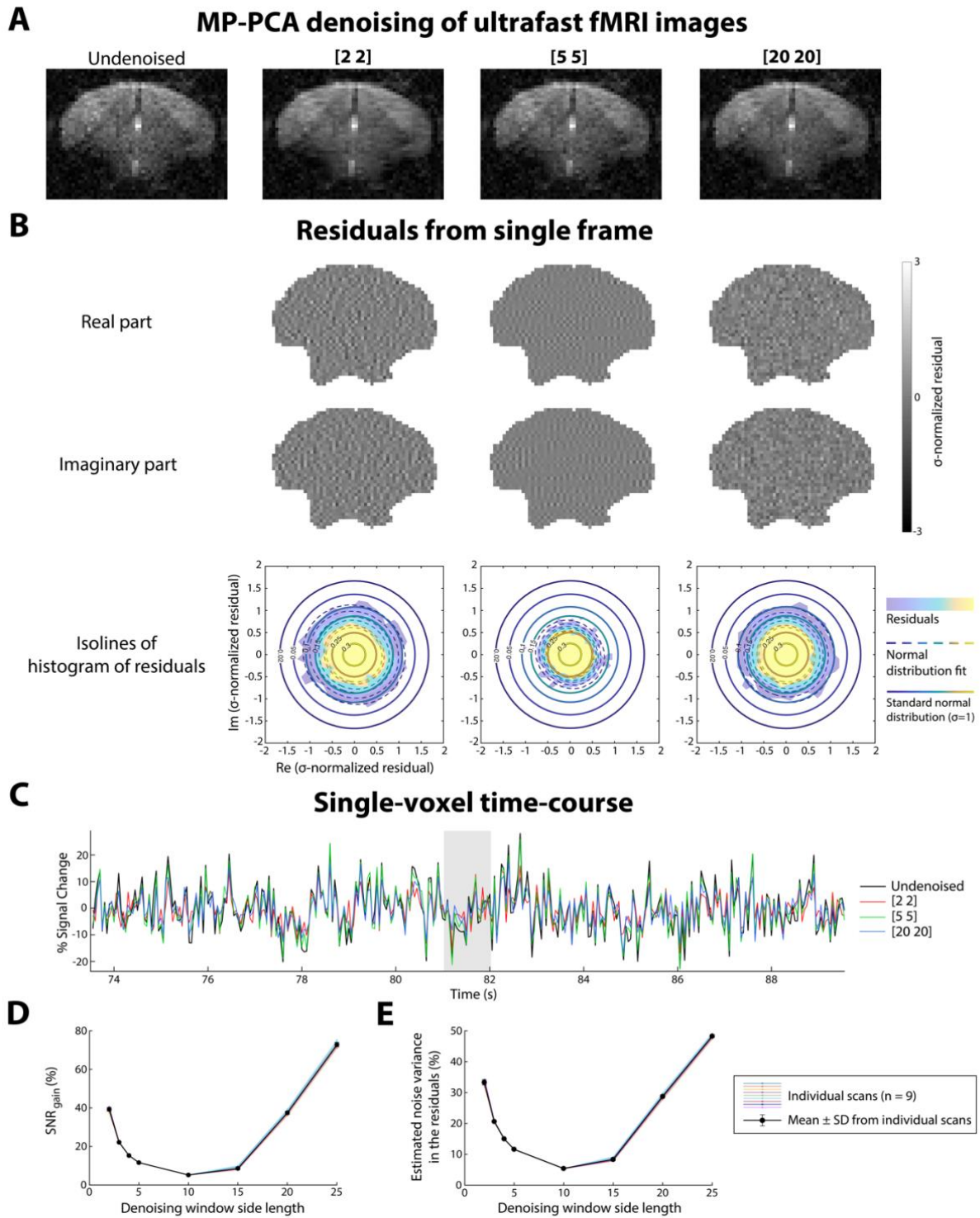

**Figure 5 – Residuals, single-voxel time-courses and SNR gain of ultrafast fMRI data after MP-PCA denoising (Strategy B). (A)** GE-EPI image of a representative ultrafast fMRI acquisition, before and after MP-PCA denoising with 3 different sliding windows ([2 2], [5 5] and [20 20]). ZF was removed from k-space prior to data denoising. **(B)** Maps of real (*top*) and imaginary

(*middle*) part of σ-normalized residuals of a single frame obtained after MP-PCA denoising with these 3 windows, together with filled contour plots containing isolines of the histogram of those residuals (*bottom*). Dashed lines represent the approximated normal distribution for the residuals whereas solid lines represent the standard normal distribution (unitary variance). **(C)** Single-voxel detrended time-courses (only repetitions 1471 to 1791 are shown) before (black line) and after MP-PCA denoising with 3 different sliding windows ([2 2] in red, [5 5] in green and [20 20] in blue). The grey area represents the second period of visual stimulation. **(D)** Average brain SNR gain across 8 different sliding windows and n = 9 different scans (represented by each color plot) of ultrafast fMRI data. The black dots with error bars and black lines represent the mean ± SD results obtained from the individual scans. **(E)** Percentage of estimated noise variance explained by the residuals across the same sliding windows and scans.

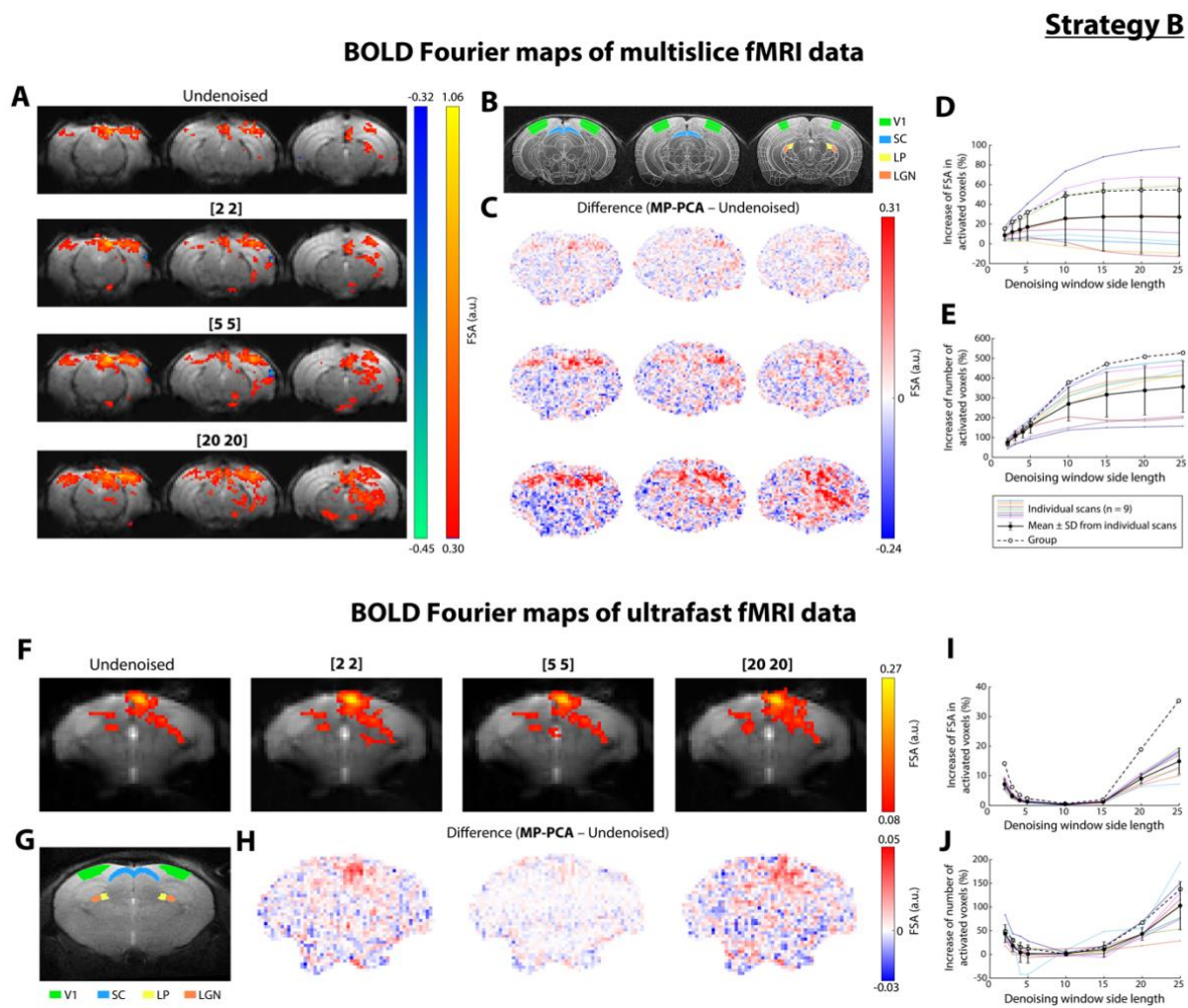

Figure 6 – Individual BOLD Fourier maps of multislice and ultrafast fMRI data after MP-PCA denoising (Strategy B). **(A)** BOLD Fourier maps of a representative multislice fMRI acquisition (3 out of 10 slices shown) obtained before and after MP-PCA denoising at the slice level with 3 different sliding windows ([2 2], [5 5] and [20 20]). ZF was removed from k-space prior to data denoising. Maps are thresholded with a minimum FSA at paradigm's fundamental frequency = 0.3 and a minimum cluster size = 10. **(B)** ROIs of the mouse visual pathway from the Allen Reference Atlas delineated on anatomical images. **(C)** Difference between the functional maps shown in (A). **(D)** Percentage increase of FSA at paradigm's fundamental

frequency in activated voxels relative to the undenoised data results across 8 different sliding windows and n = 9 different scans (represented by each color plot) of multislice fMRI data. **(E)** Percentage increase of number of activated voxels in the BOLD maps relative to the undenoised data results across the same sliding windows and scans. The empty dots and dashed lines in (D) and (E) are the results from the group BOLD maps. **(F-H)** Same as (A-C), respectively, but for a representative ultrafast fMRI scan. Maps are thresholded with a minimum sum of FSA at paradigm's fundamental frequency and two following harmonics = 0.08 and a minimum cluster size = 8. **(I-J)** Same as (D-E), respectively, but for n = 9 different scans of ultrafast fMRI data.

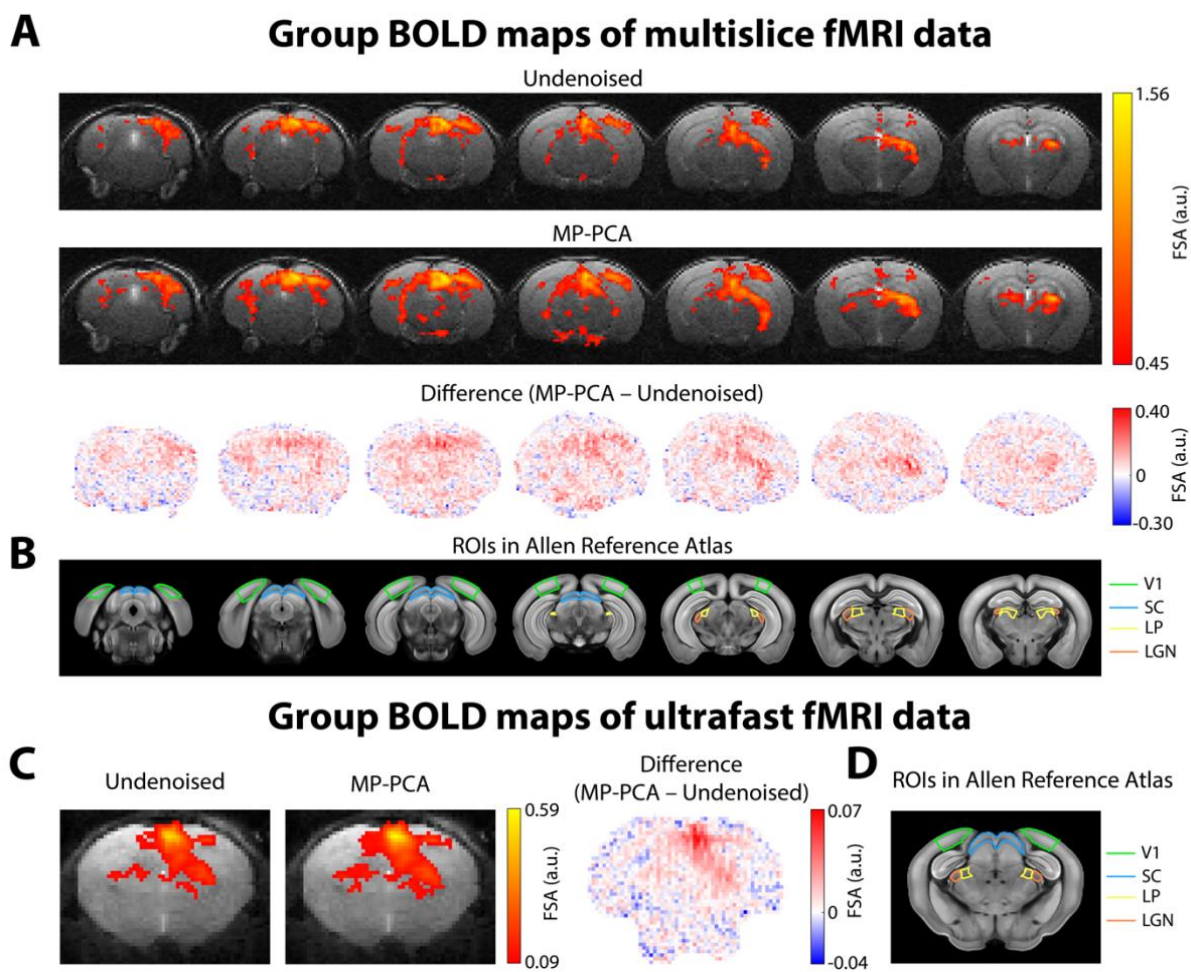

**Figure 7 – Group BOLD Fourier maps of multislice and ultrafast fMRI data after MP-PCA denoising (Strategy B). (A)** Group BOLD Fourier maps (n = 9 scans) of multislice fMRI data (7 out of 10 slices shown) obtained before (*top*) and after (*middle*) MP-PCA denoising at the slice level with a sliding window = [2 2], and respective difference map (*bottom*). Maps are thresholded with a minimum FSA at paradigm's fundamental frequency = 0.45 and a minimum cluster size = 10, and displayed above anatomical images. **(B)** ROIs of the mouse visual pathway delineated on the Allen Reference Atlas. **(C-D)** Same as (A-B), respectively, but for ultrafast fMRI data (n = 9 scans). Maps are thresholded with a minimum sum of FSA at paradigm's fundamental frequency and two following harmonics = 0.09 and a minimum cluster size = 8.

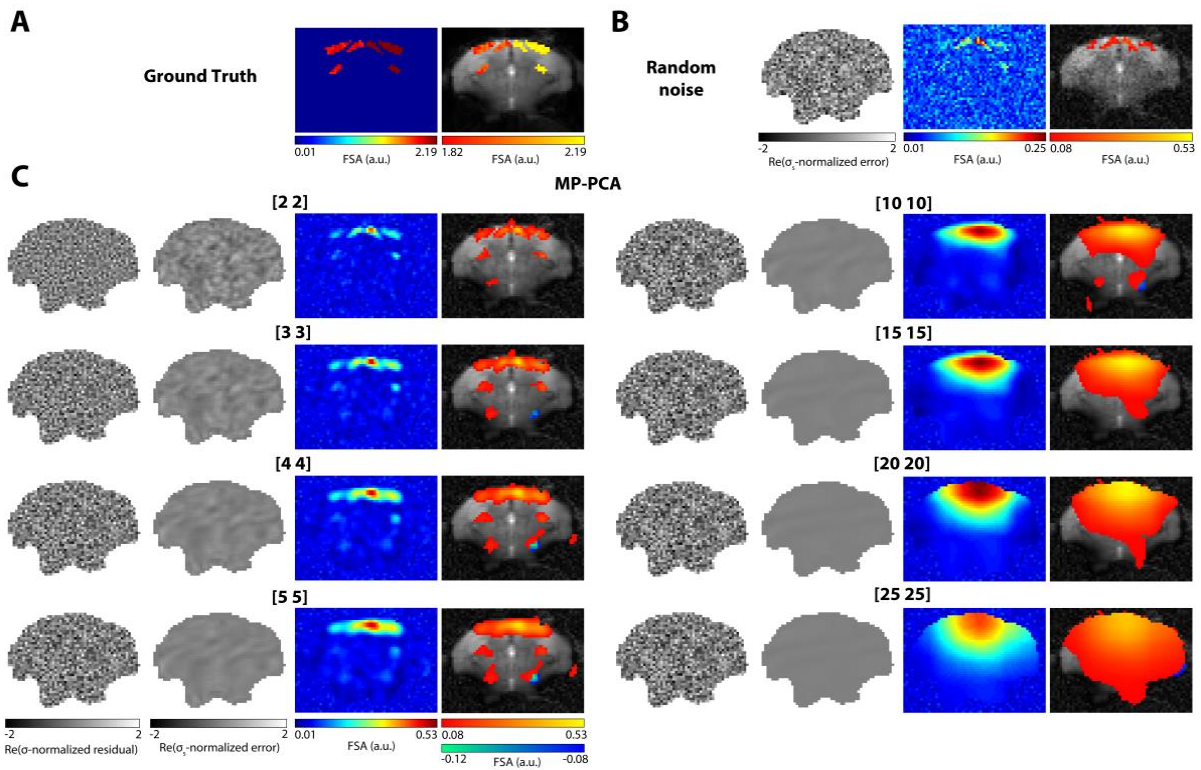

**Figure 8 – Residual, error and BOLD maps of simulated ultrafast fMRI data (with tSNR = 6.7 and 0.75-2.25% BOLD changes) after MP-PCA denoising. (A)** BOLD Fourier map of ultrafast fMRI data simulated without any source of noise, with (*right*) and without (*left*) threshold. **(B)** Same as (A) after addition of Gaussian white noise so that the average tSNR in the brain = 6.7. A map of $\sigma_s$-normalized error of a single frame is displayed on the left. **(C)** Maps of $\sigma$-normalized residuals and $\sigma_s$-normalized error of a single frame (*left*) and BOLD Fourier maps with (*right*) and without (*middle*) threshold upon MP-PCA denoising with 8 different sliding windows. Maps are thresholded with a minimum sum of FSA at paradigm's fundamental frequency and two following harmonics = 0.08 and a minimum cluster size = 8. A representative image of the data is shown below the maps.

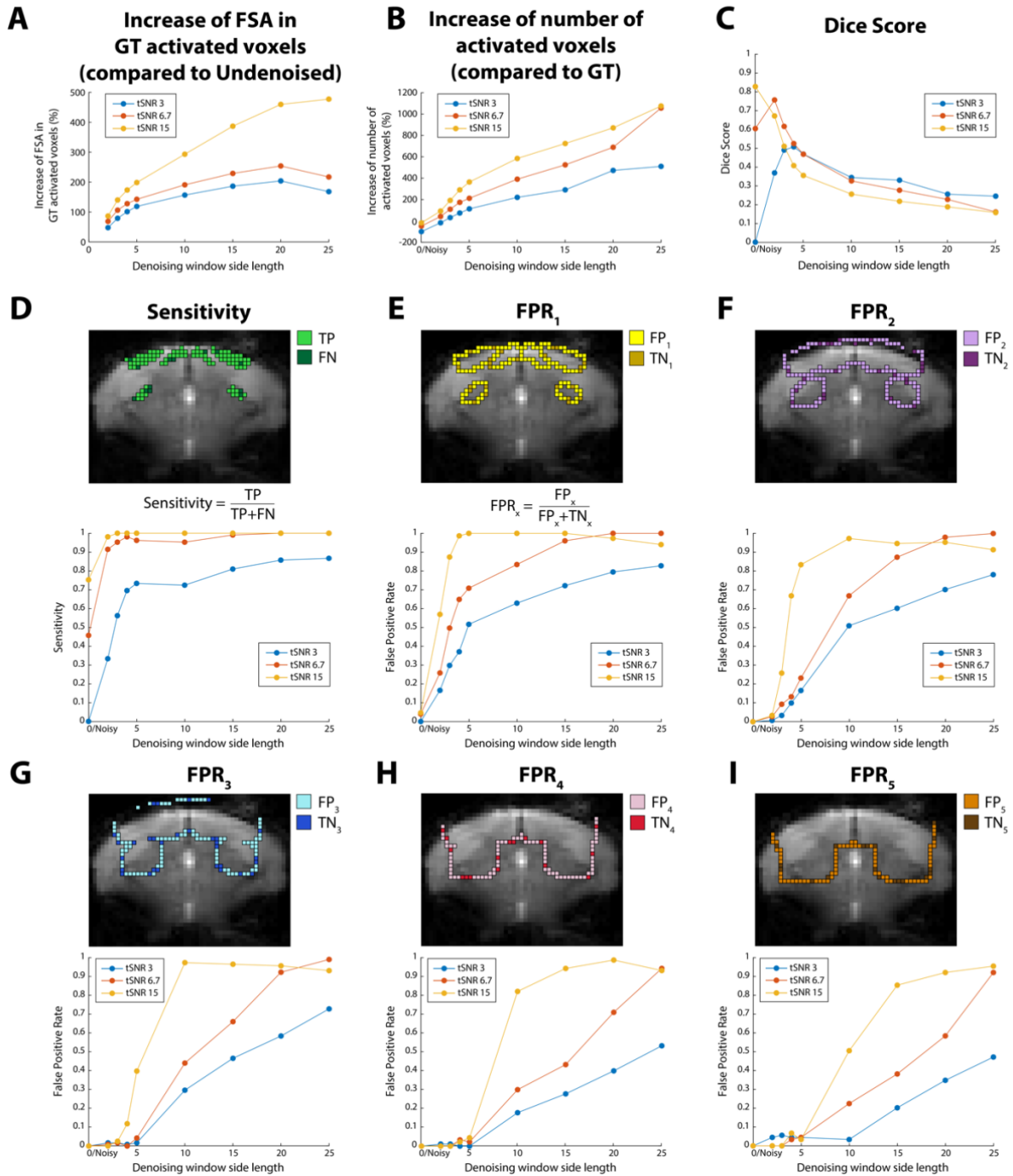

Figure 9 – Key metrics of BOLD maps of simulated ultrafast fMRI data (with 3 different tSNRs) after MP-PCA denoising. **(A)** Percentage increase of FSA at paradigm's fundamental frequency and two following harmonics in GT activated voxels (i.e., the voxels above the threshold in the BOLD map obtained from GT data) relative to the undenoised data results across 8 different sliding windows and 3 different tSNRs (3 in blue, 6.7 in orange, 15 in yellow) of simulated ultrafast fMRI data. **(B)** Percentage increase of number of activated voxels in the BOLD maps relative to the GT data results across different sliding windows and tSNRs. **(C)** Dice score of activation maps across different sliding windows and tSNRs. **(D)** (*Top*) The sensitivity of activation maps is defined as the ratio between the number of true positive (TP) voxels and the number of true positive and false negative (FN) voxels. (*Bottom*) Sensitivity of maps across

different sliding windows and tSNRs. **(E-I)** (*Top*) The FPR is defined as the number of false positive (FP) voxels over the number of false positive and true negative (TN) voxels within the 1- to 5-pixel perimeter layers of the GT activation volume inside the brain mask (FPR$_1$ to FPR$_5$). (*Bottom*) FPR$_1$ to FPR$_5$ of maps across different sliding windows and tSNRs.

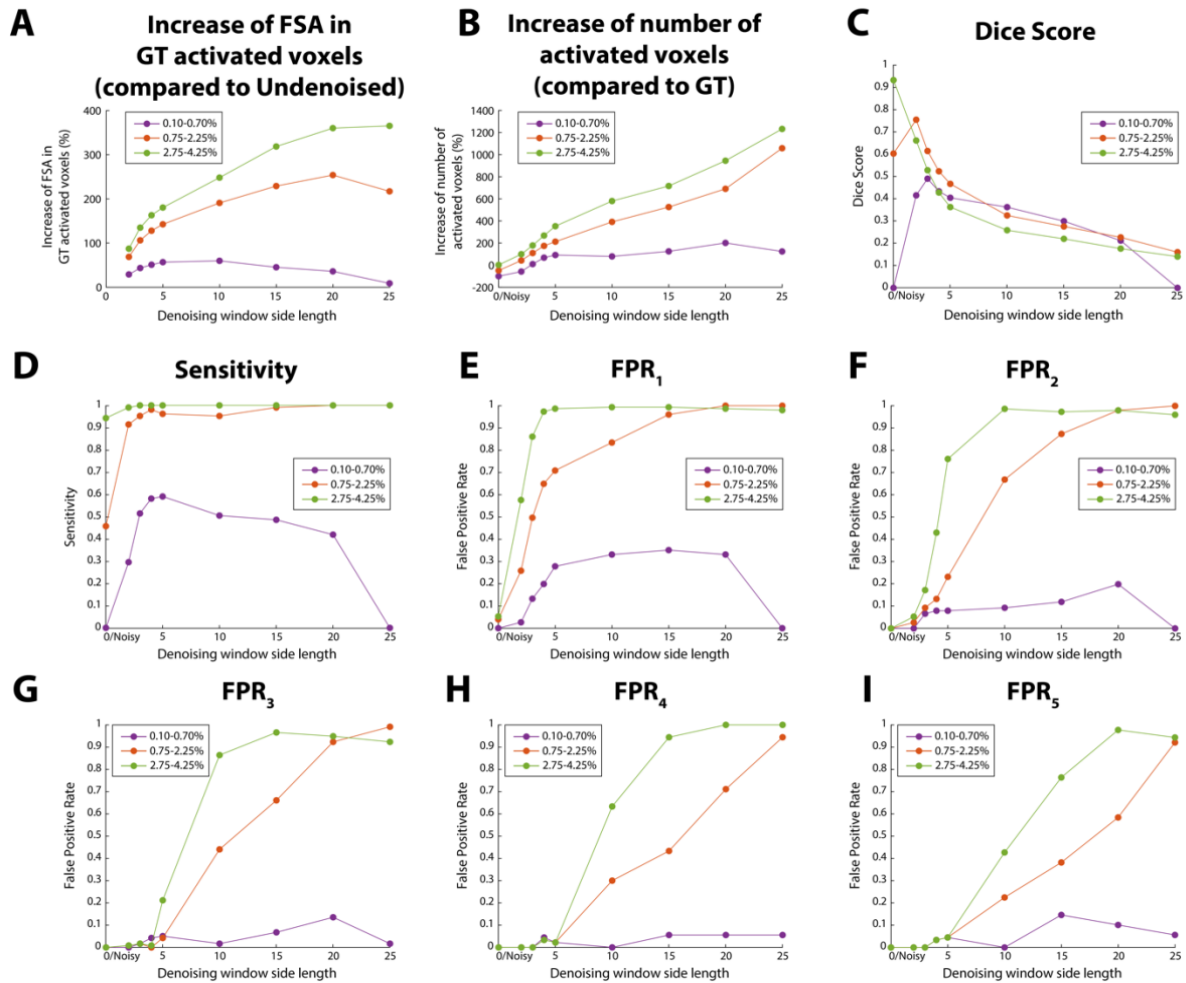

**Figure 10 – Key metrics of BOLD maps of simulated ultrafast fMRI data (with 3 different intervals of % BOLD change) after MP-PCA denoising. (A)** Percentage increase of FSA at paradigm's fundamental frequency and two following harmonics in GT activated voxels (i.e., the voxels above the threshold in the BOLD map obtained from GT data) relative to the undenoised data results across 8 different sliding windows and 3 different intervals of % BOLD change (0.10-0.70% in purple, 0.75-2.25% in orange, 2.75-4.25% in green) of simulated ultrafast fMRI data. **(B)** Percentage increase of number of activated voxels in the BOLD maps relative to the GT data results, and **(C)** Dice score, **(D)** Sensitivity and **(E-I)** FPR$_1$ to FPR$_5$ of activation maps across different sliding windows and intervals of % BOLD change.

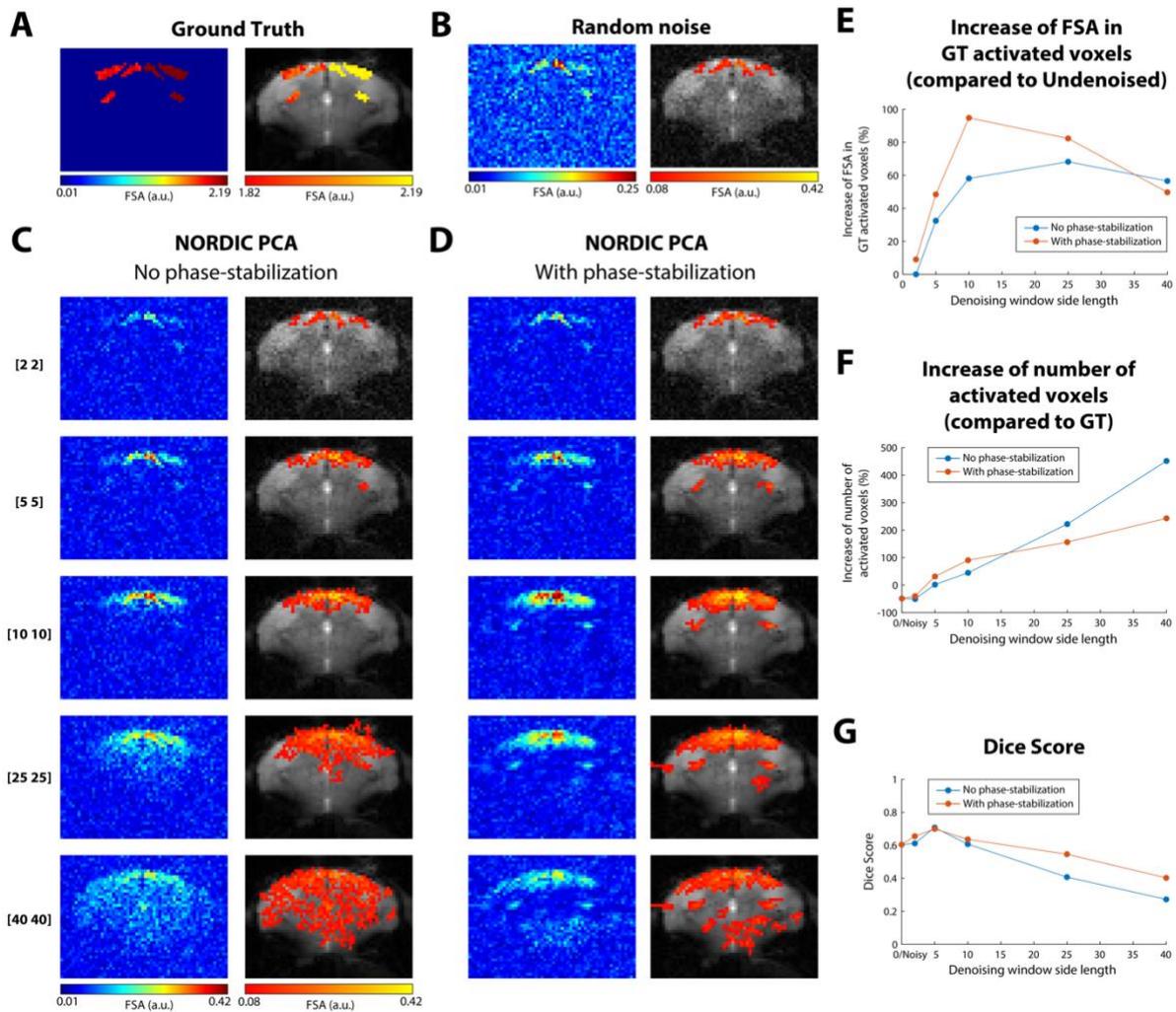

**Figure 11 – BOLD maps of simulated ultrafast fMRI data (tSNR = 6.7) after NORDIC PCA denoising. (A)** BOLD Fourier map of ultrafast fMRI data simulated without any source of noise, with (*right*) and without (*left*) threshold. **(B)** Same as (A) after addition of Gaussian white noise so that the average tSNR in the brain = 6.7. **(C)** BOLD Fourier maps with (*right*) and without (*left*) threshold upon NORDIC PCA denoising with 5 different sliding windows and no phase-stabilization correction. Maps are thresholded with a minimum sum of FSA at paradigm's fundamental frequency and two following harmonics = 0.08 and a minimum cluster size = 8. A representative image of the data is shown below the maps. **(D)** Same as (C) but after NORDIC PCA denoising with phase-stabilization correction. **(E)** Percentage increase of FSA at paradigm's fundamental frequency and two following harmonics in GT activated voxels (i.e., the voxels above the threshold in the BOLD map obtained from GT data) relative to the undenoised data results across 5 different sliding windows. Blue datapoints: no phase-stabilization correction prior to NORDIC denoising. Orange datapoints: phase-stabilization correction applied before NORDIC denoising. **(F)** Percentage increase of number of activated voxels in the BOLD maps relative to the GT data results across different sliding windows. **(G)** Dice score of activation maps across different sliding windows.

# Supplementary Figures

**Video S1 – Videos of multislice and ultrafast fMRI data upon MP-PCA denoising after ZF removal from k-space (Strategy B). (A)** Videos of GE-EPI images from a representative slice of a multislice fMRI acquisition, before and after MP-PCA denoising at the slice level with 8 different sliding windows. ZF was removed from k-space prior to data denoising. The 340 acquired repetitions are shown. **(B)** Same as (A) but for the single slice of a representative ultrafast fMRI scan. Only 340 out of 9000 acquired repetitions are shown, specifically, from repetition 641 to 980.

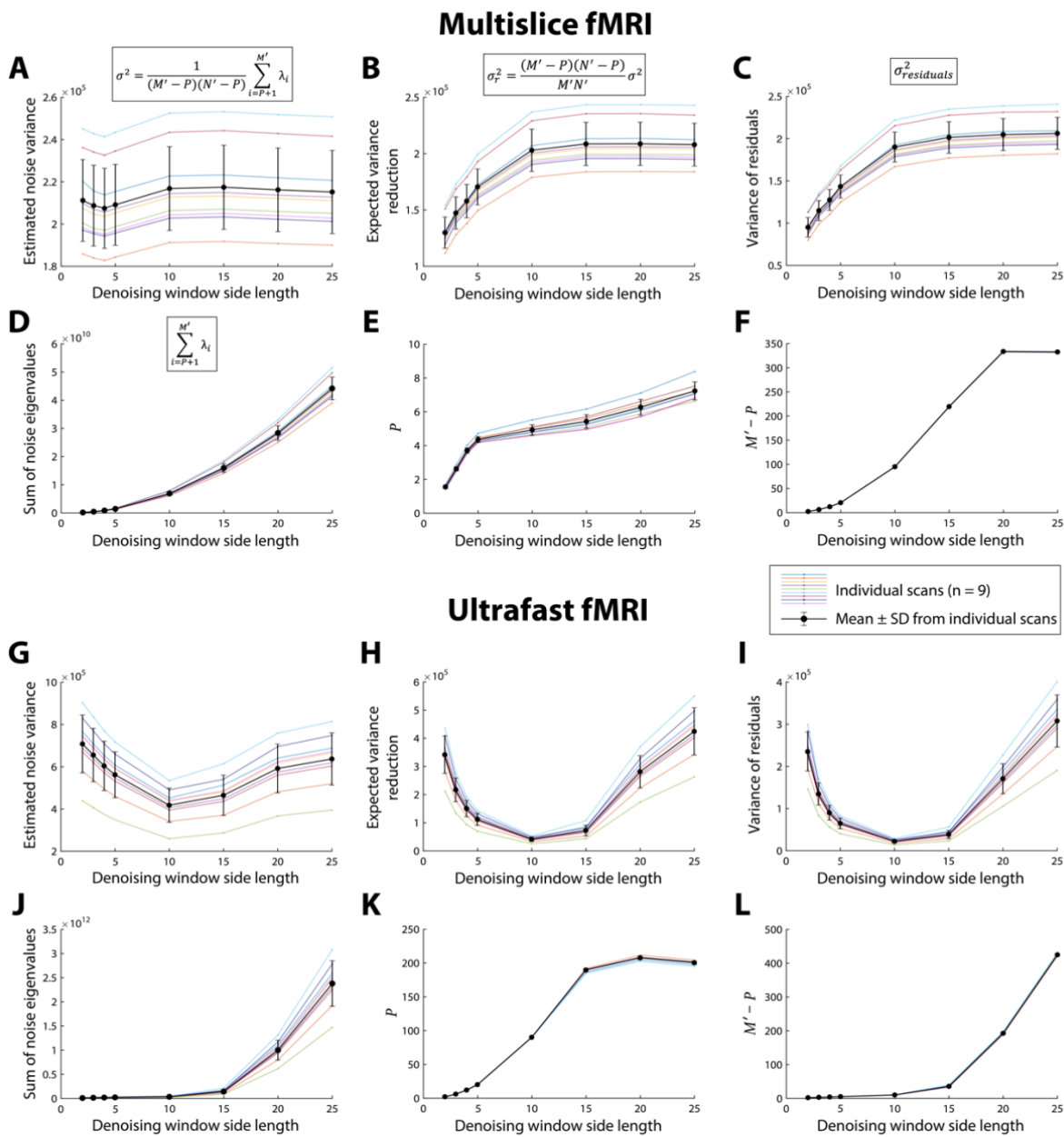

**Figure S1 – MP-PCA denoising output metrics for multislice and ultrafast fMRI data (Strategy B). (A)** Average brain estimated noise variance $\sigma^2$ across 8 different sliding windows and n = 9 different scans (represented by each color plot) of multislice fMRI data. The black dots with error bars and black lines represent the mean ± SD results obtained from the individual scans. Average brain **(B)** expected variance reduction $\sigma_r^2$ obtained by truncating Gaussian noise components only, **(C)** variance of (unnormalized) residuals $\sigma_{residuals}^2$, **(D)** sum of noise eigenvalues, **(E)** number of retained $P$ signal components and **(F)** number of eliminated "noise" components $M' - P$ across the same sliding windows and scans. **(G-L)** Same as (A-F), respectively, but for n = 9 different scans of ultrafast fMRI data.

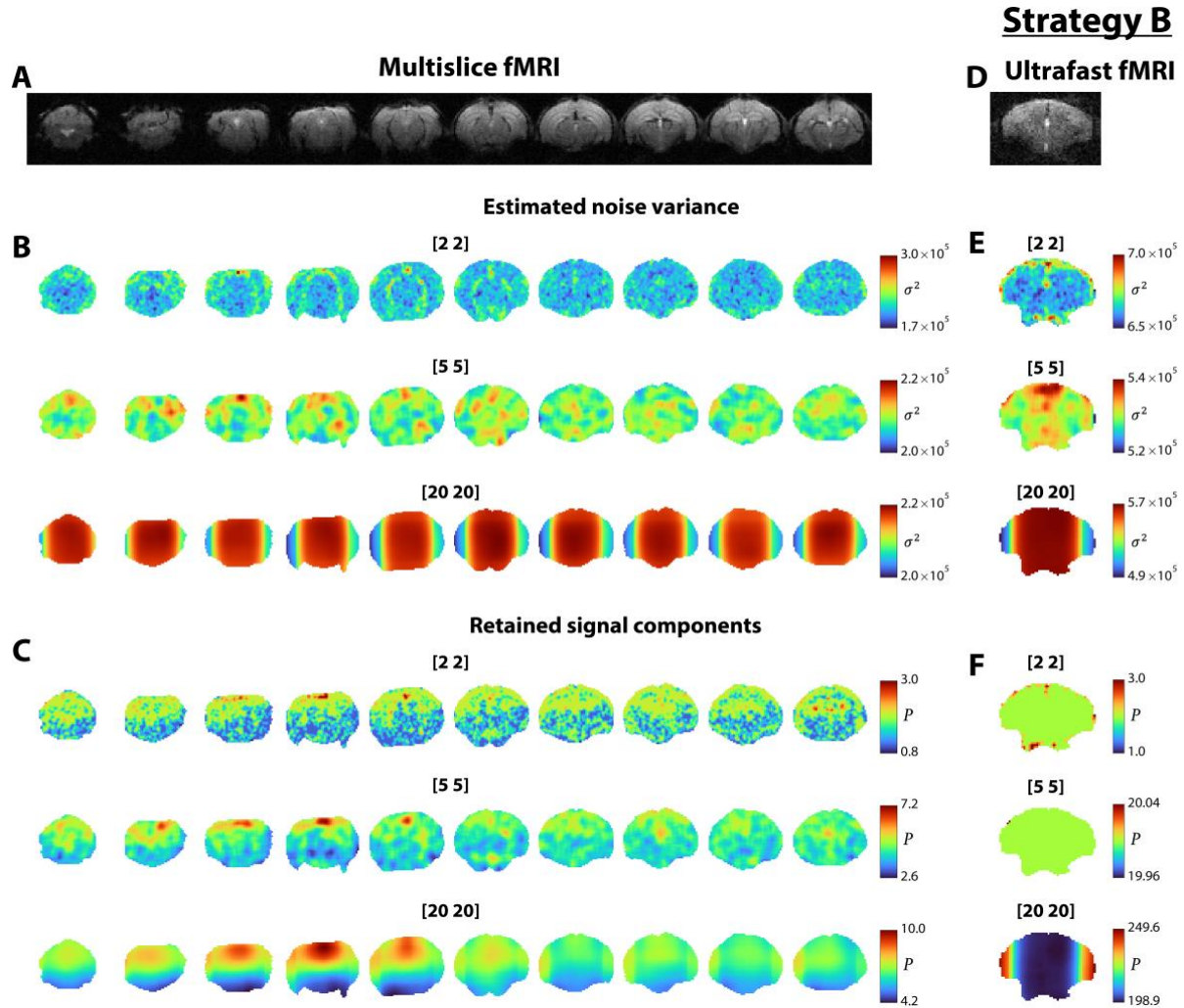

**Figure S2 – MP-PCA denoising output metrics across brain for multislice and ultrafast fMRI data (Strategy B). (A)** Single GE-EPI image obtained from a representative multislice fMRI acquisition. **(B)** Estimated noise variance $\sigma^2$ maps obtained after MP-PCA denoising of a multislice fMRI dataset with 3 different sliding windows ([2 2], [5 5] and [20 20]) out of the 8 used in this study. **(C)** Number of $P$ signal components retained after MP-PCA denoising of multislice data with these 3 windows. **(D-E)** Same as (A-C), respectively, but for a representative ultrafast fMRI scan.

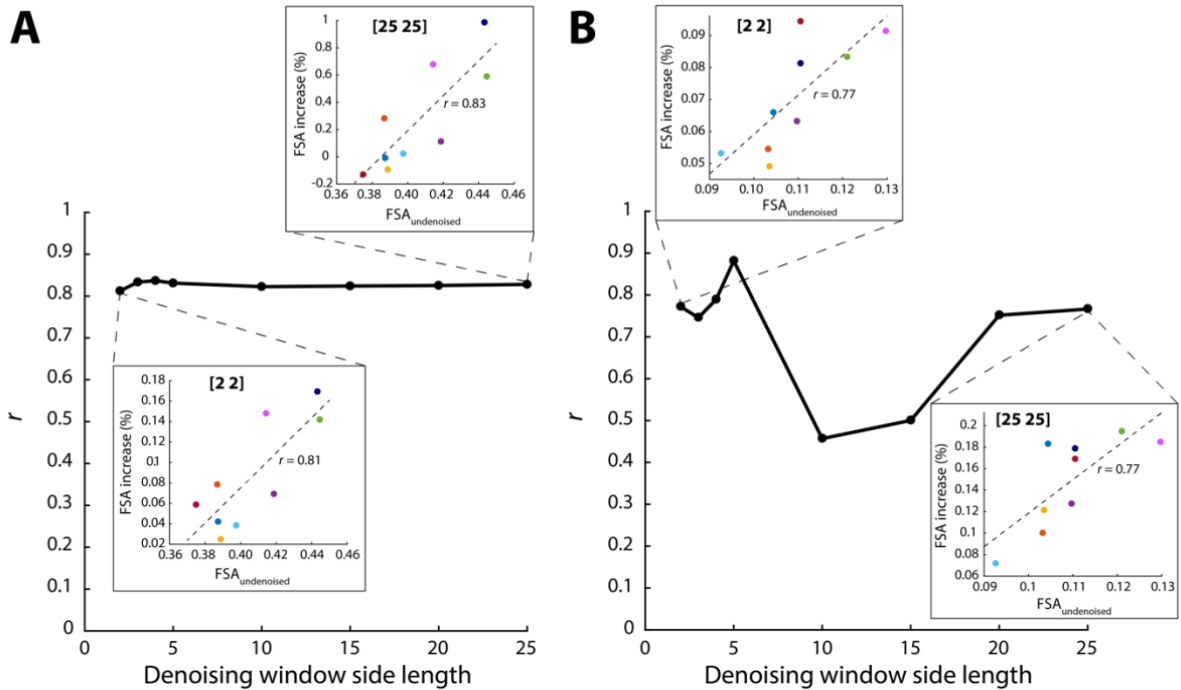

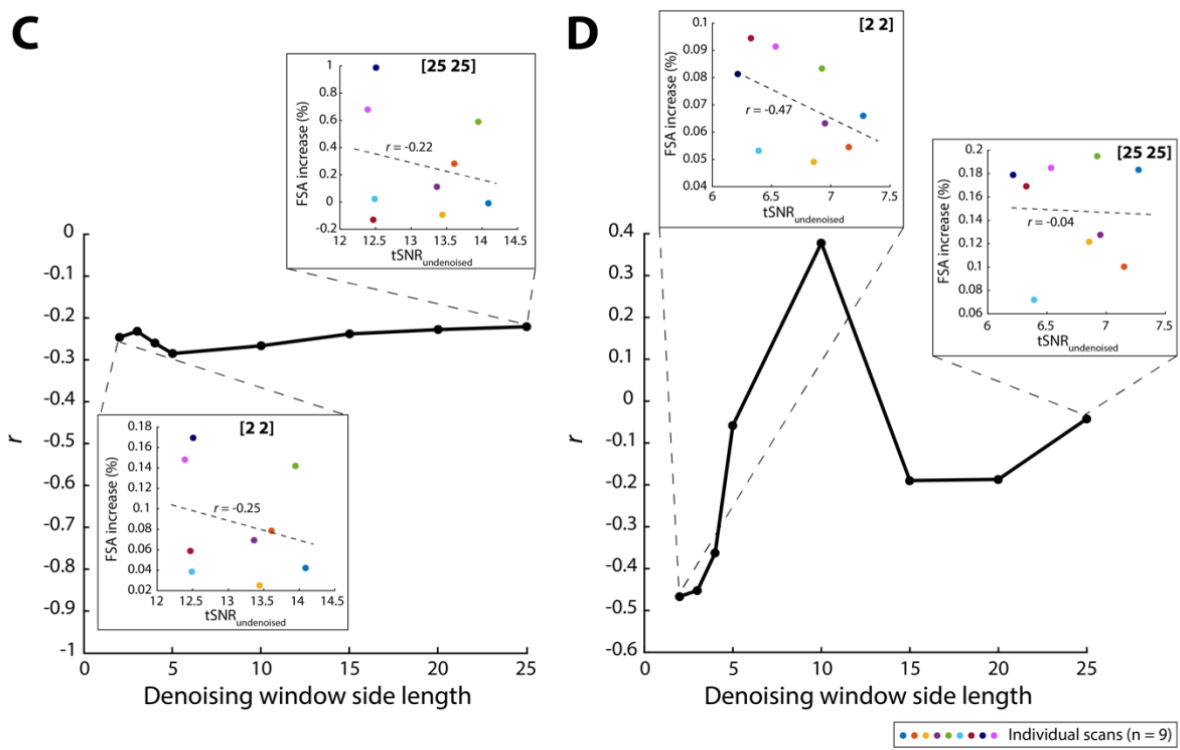

**Figure S3 – Pearson's correlation values between initial FSA or tSNR values and FSA increase values of different individual scans. (A)** Correlation between the average FSA extracted from activated voxels of undenoised data BOLD maps and the values of FSA increase of n = 9 different individual multislice fMRI scans, for 8 different sliding windows. The two insets show

the scatter plots of those variables for the smallest ([2 2]) and largest ([25 25]) sliding windows used in this study. **(B)** Same as (A) but for n = 9 individual ultrafast fMRI scans. **(C-D)** Same as (A-B) but for the correlation between tSNR values estimated from undenoised data and the values of FSA increase for each sliding window.

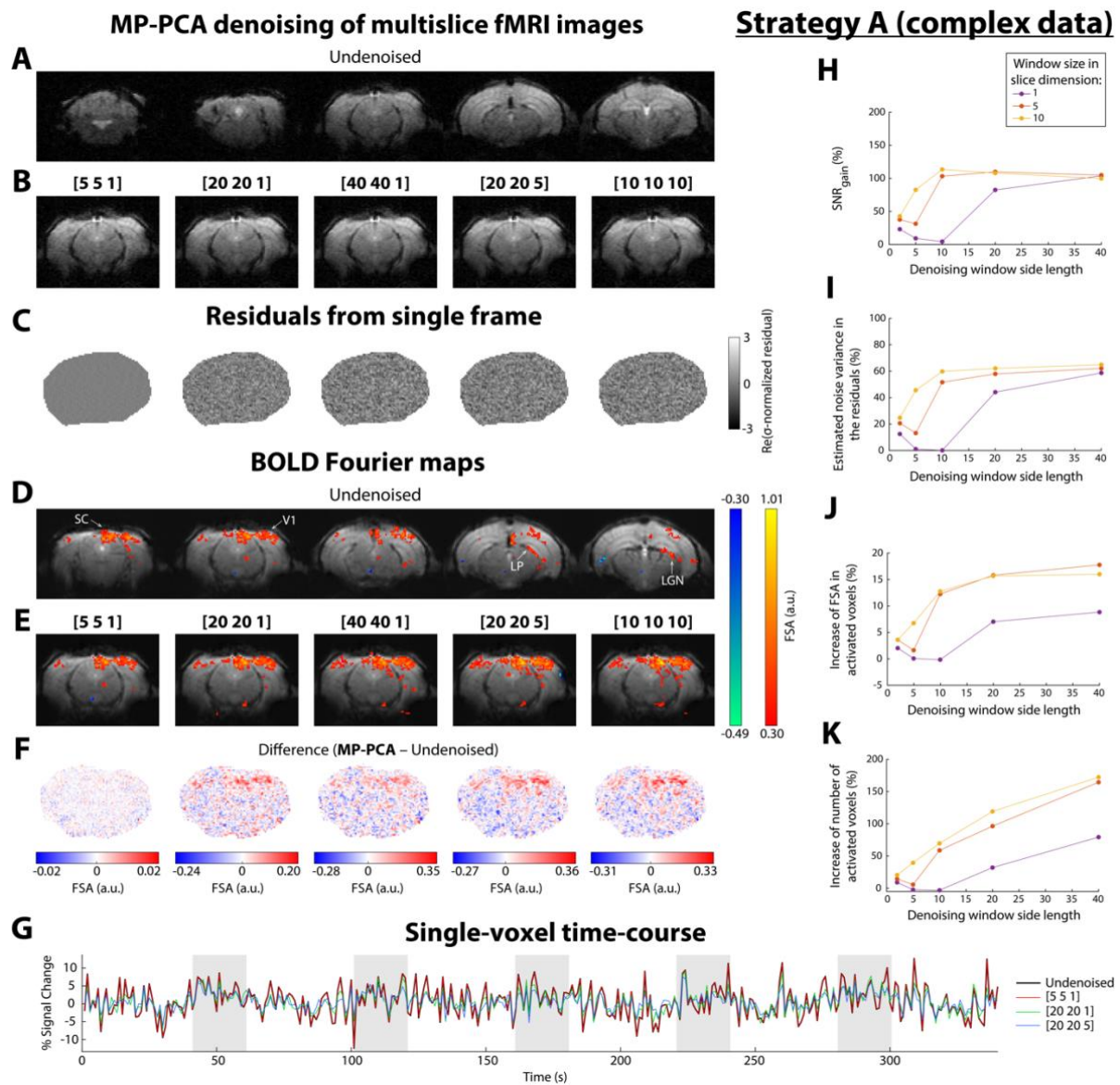

**Figure S4 – MP-PCA denoising of complex multislice fMRI data reconstructed by vendor software (Strategy A with complex data). (A)** Single GE-EPI image obtained from a representative multislice fMRI acquisition, before (5 out of 10 slices shown) and **(B)** after (1 out of 10 slices shown) application of MP-PCA denoising on complex data with 5 different sliding windows ([5 5 1], [20 20 1], [40 40 1], [20 20 5] and [10 10 10]) immediately after outlier correction. **(C)** Map of real part of $\sigma$-normalized residuals of a single frame obtained after MP-PCA denoising with these 5 windows. **(D)** BOLD Fourier maps obtained before and **(E)** after MP-PCA denoising with these 5 windows. Maps are thresholded with a minimum FSA at paradigm's fundamental frequency = 0.3 and a minimum cluster size = 10. **(F)** Difference between the functional maps shown in (E) and (D). **(G)** Single-voxel detrended time-courses before (black line) and after MP-PCA denoising with 3 different sliding windows ([5 5 1] in red, [20 20 1] in green and [20 20 5] in blue). Grey areas represent the periods of visual stimulation.

**(H)** Average brain SNR gain, **(I)** percentage of estimated noise variance explained by the residuals, **(J)** percentage increase of FSA in activated voxels, and **(K)** increase in spatial extent of activation, obtained after MP-PCA denoising with 15 different sliding windows.

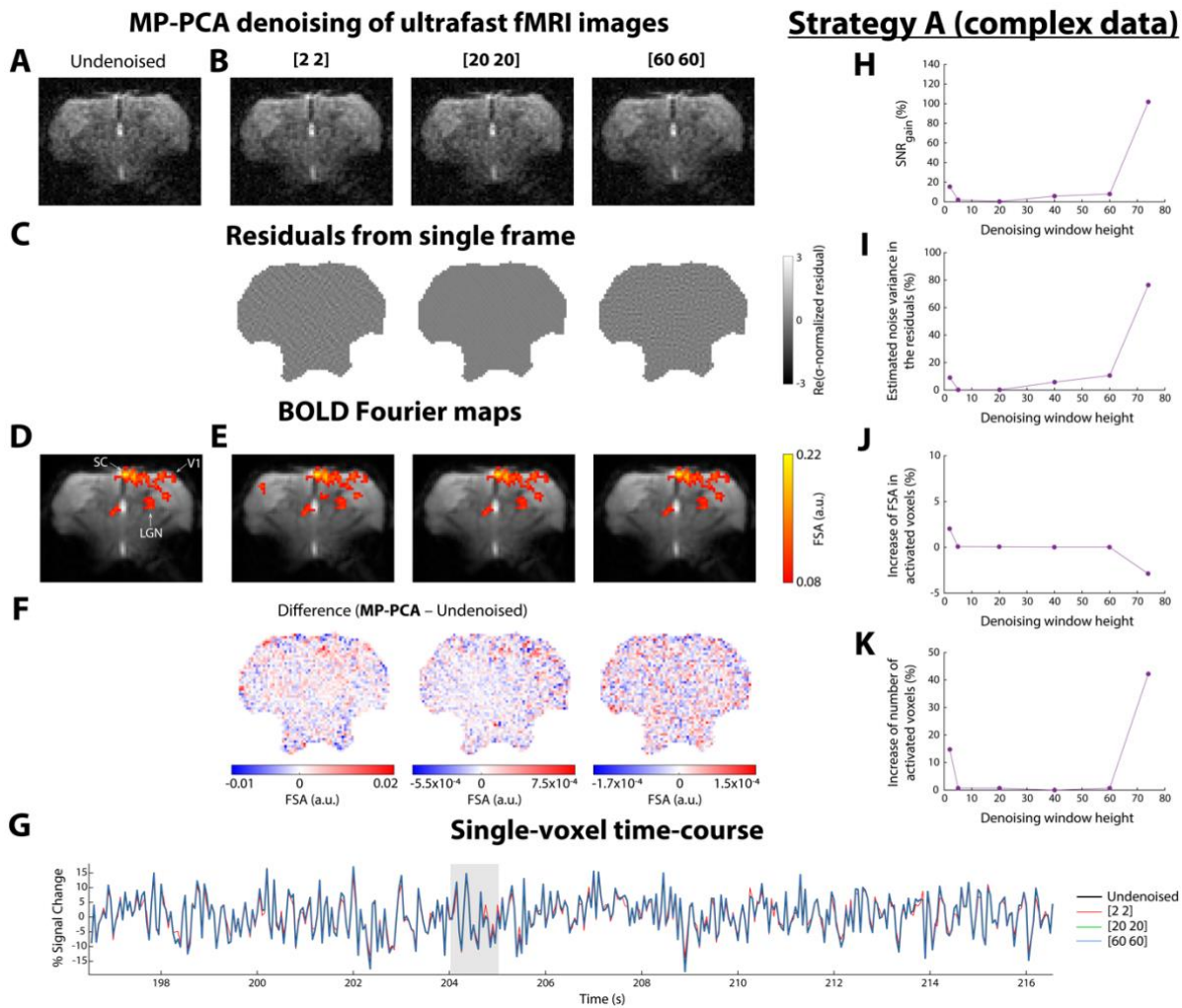

**Figure S5 – MP-PCA denoising of complex ultrafast fMRI data reconstructed by vendor software (Strategy A with complex data). (A)** GE-EPI image obtained from a representative ultrafast fMRI acquisition, before and **(B)** after application of MP-PCA denoising on complex data with 3 different sliding windows ([2 2], [20 20] and [60 60]) immediately after outlier correction. **(C)** Map of real part of $\sigma$-normalized residuals of a single frame obtained after MP-PCA denoising with these 3 windows. **(D)** BOLD Fourier maps obtained before and **(E)** after MP-PCA denoising with these 3 windows. Maps are thresholded with a minimum sum of FSA at paradigm's fundamental frequency and two following harmonics = 0.08 and a minimum cluster size = 8. **(F)** Difference between the functional maps shown in (E) and (D). **(G)** Single-voxel detrended time-courses (only repetitions 3931 to 4331 are shown) before (black line) and after MP-PCA denoising with 3 different sliding windows ([2 2] in red, [20 20] in green and [60 60] in blue). The grey area represents the fifth period of visual stimulation. **(H)** Average brain SNR gain, **(I)** percentage of estimated noise variance explained by the residuals, **(J)** percentage increase of FSA in activated voxels, and **(K)** increase in spatial extent of activation, obtained after MP-PCA denoising with 6 different sliding windows.

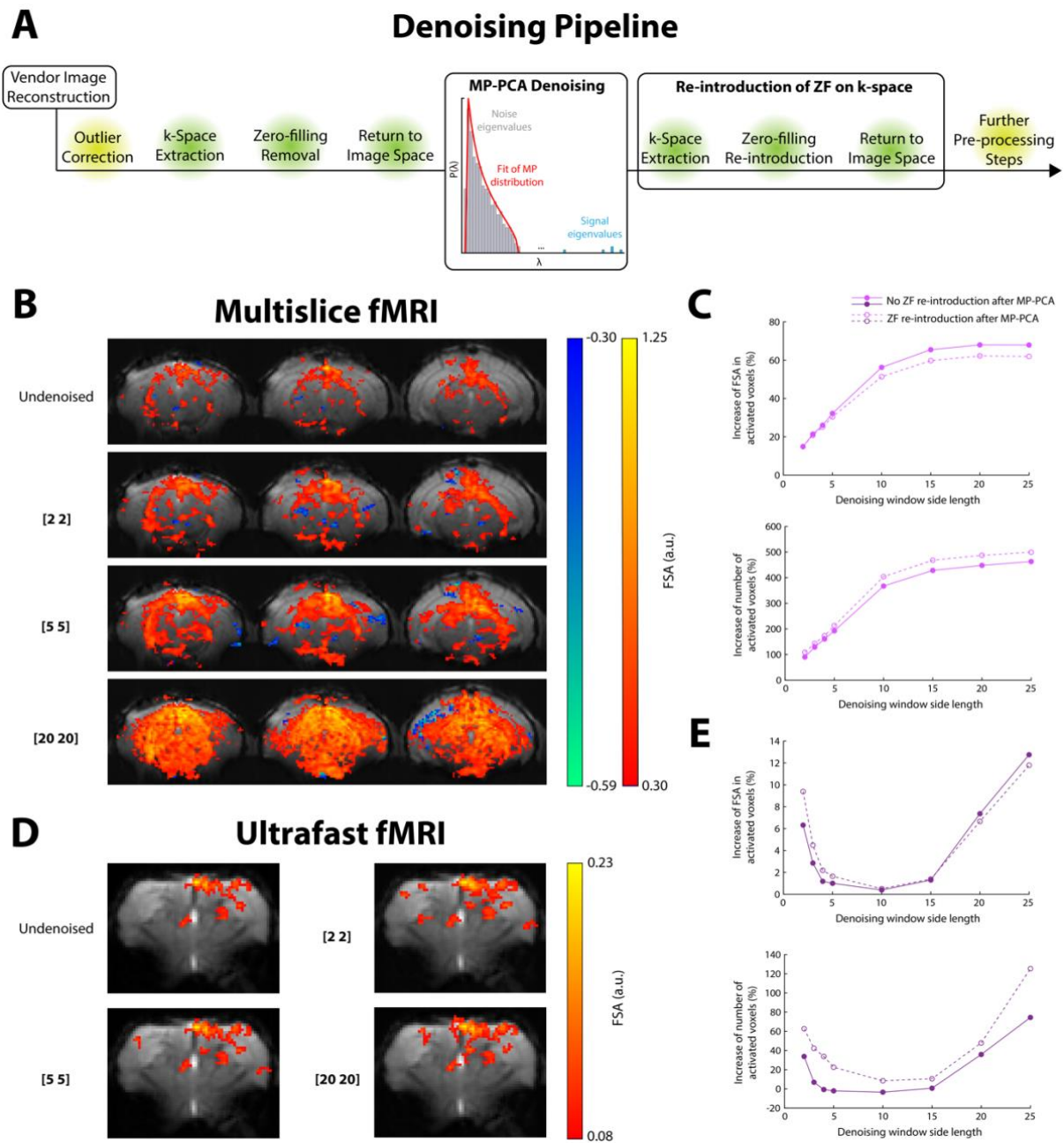

**Figure S6 – BOLD Fourier maps of multislice and ultrafast fMRI data after MP-PCA denoising and re-introduction of ZF regions on k-space. (A)** Denoising flowchart. ZF was removed from k-space prior to data denoising and re-introduced once denoising had finished. **(B)** BOLD Fourier maps of a representative multislice fMRI acquisition obtained before and after MP-PCA denoising with 3 different sliding windows ([2 2], [5 5] and [20 20]) out of the 8 used in this study. Maps are thresholded with a minimum FSA at paradigm's fundamental frequency = 0.3 and a minimum cluster size = 10. **(C)** Percentage increase of FSA in activated voxels (*top*) and percentage increase of number of activated voxels in the BOLD maps (*bottom*) relative to the undenoised data results across the 8 sliding windows in this multislice fMRI scan. Empty dots with dashed lines and filled dots with solid lines represent the variations observed in maps obtained from data with and without re-introduction of ZF regions on k-space after denoising, respectively. **(D-E)** Same as (B-C), respectively, but for a representative ultrafast fMRI scan. Maps are thresholded with a minimum sum of FSA at paradigm's fundamental frequency and two following harmonics = 0.08 and a minimum cluster size = 8.

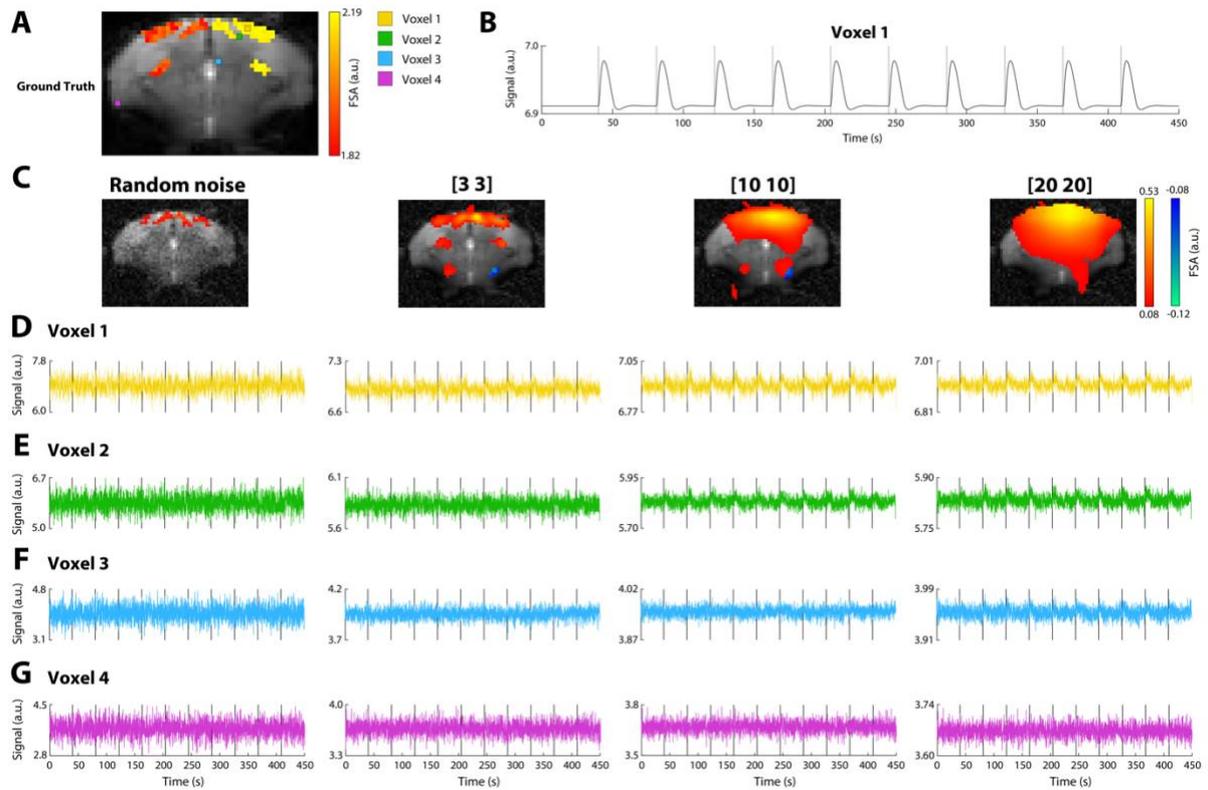

**Figure S7 – Single-voxel time-courses of simulated ultrafast fMRI data (with tSNR = 6.7 and 0.75-2.25% BOLD changes) after MP-PCA denoising. (A)** Voxels from which time-courses were extracted defined above the BOLD Fourier map obtained from GT data: voxel 1 is located above the (activated) right V1 whereas voxels 2 to 4 (non-activated) are gradually further away from activated regions. **(B)** Time-course of voxel 1 before addition of noise. **(C)** BOLD Fourier maps of simulated ultrafast fMRI data after addition of Gaussian white noise (so that the average tSNR in the brain = 6.7) and after MP-PCA denoising with 3 different sliding windows ([3 3], [10 10] and [20 20]). **(D-G)** Time-courses of voxels 1 to 4, respectively, after addition of noise and after MP-PCA denoising with these 3 windows. Black lines represent the periods of visual stimulation. For visualization purposes, time-courses were smoothed with a moving average filter of span = 5.

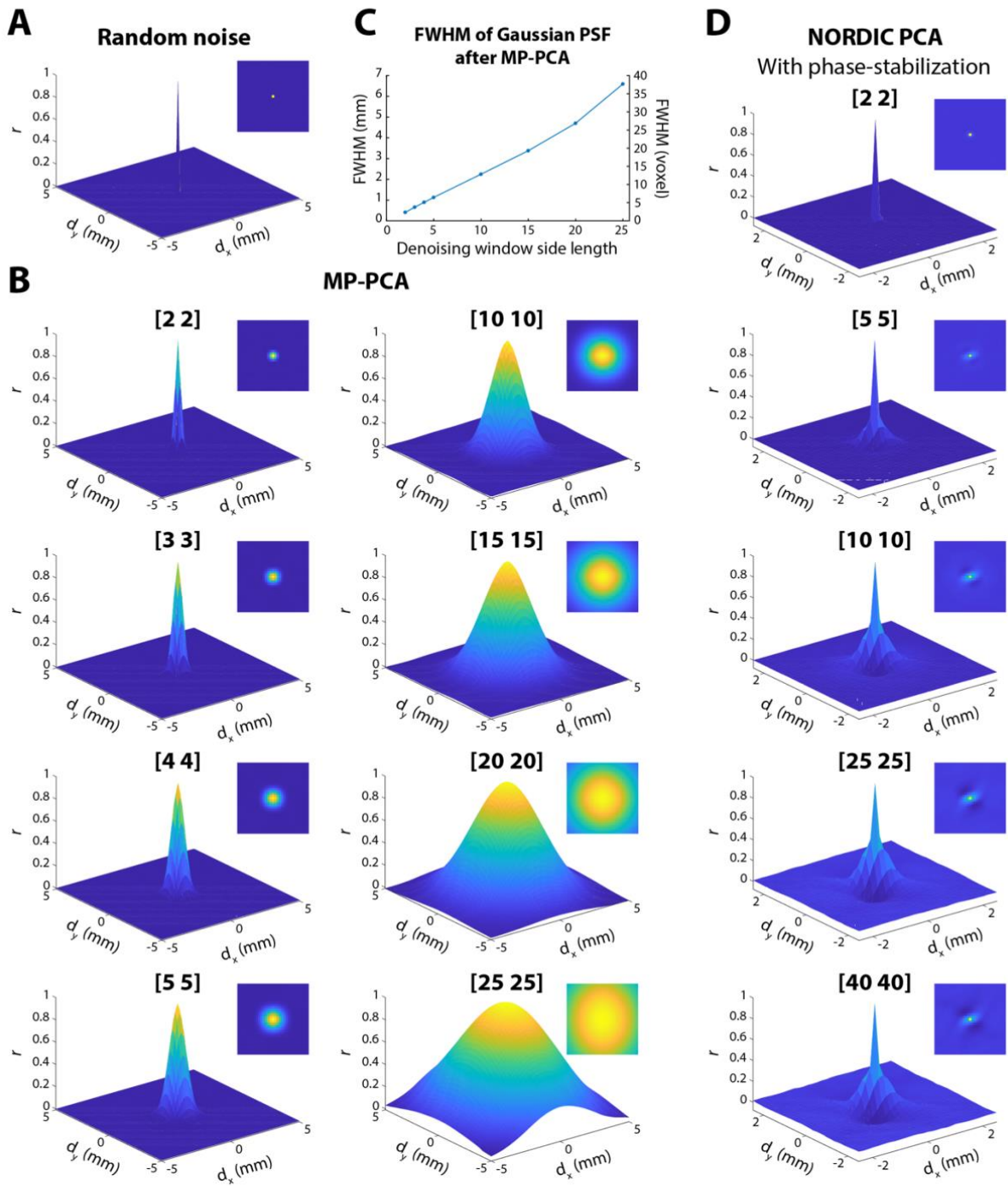

**Figure S8 – Noise kernel of simulated ultrafast fMRI data (with tSNR = 6.7 and 0.75-2.25% BOLD changes) before and after MP-PCA or NORDIC denoising. (A)** Noise kernel estimated for undenoised ultrafast data. When signal is uncorrelated between voxels, the kernel shows a narrow spike in the center. **(B)** Noise kernels obtained after MP-PCA denoising with 8 different sliding windows. **(C)** Estimated FWHM of the Gaussian PSF obtained after MP-PCA denoising across 8 sliding windows. **(D)** Noise kernels obtained after NORDIC PCA denoising (with phase-stabilization) using 5 different sliding windows. All top right insets show the noise kernel obtained on a 5 x 5 mm$^2$ range, i.e., with 31 x 31 voxels.

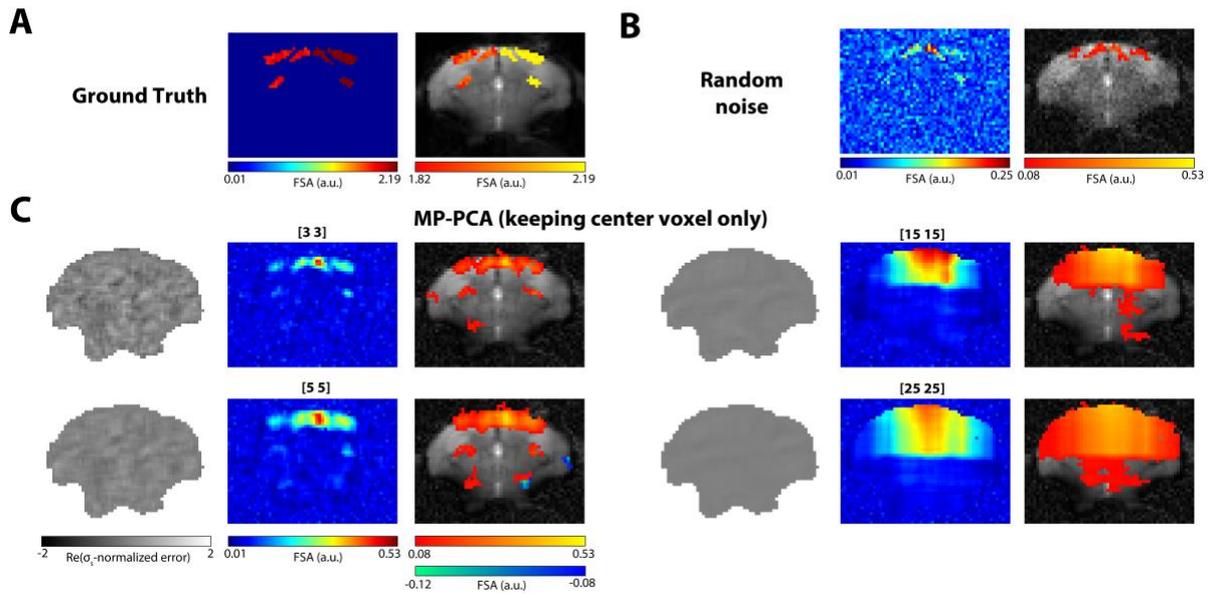

**Figure S9 – Error and BOLD maps of simulated ultrafast fMRI data (with tSNR = 6.7 and 0.75-2.25% BOLD changes) after MP-PCA denoising (only keeping the center voxel of each patch).** **(A)** BOLD Fourier map of ultrafast fMRI data simulated without any source of noise, with (*right*) and without (*left*) threshold. **(B)** Same as (A) after addition of Gaussian white noise so that the average tSNR in the brain = 6.7. **(C)** Maps of $\sigma_s$-normalized error of a single frame (*left*) and BOLD Fourier maps with (*right*) and without (*middle*) threshold upon MP-PCA denoising with 8 different sliding windows. Instead of averaging the results of overlapping voxels from multiple patches after denoising, only the center voxel of each patch was kept. Maps are thresholded with a minimum sum of FSA at paradigm's fundamental frequency and two following harmonics = 0.08 and a minimum cluster size = 8. A representative image of the data is shown below the maps.

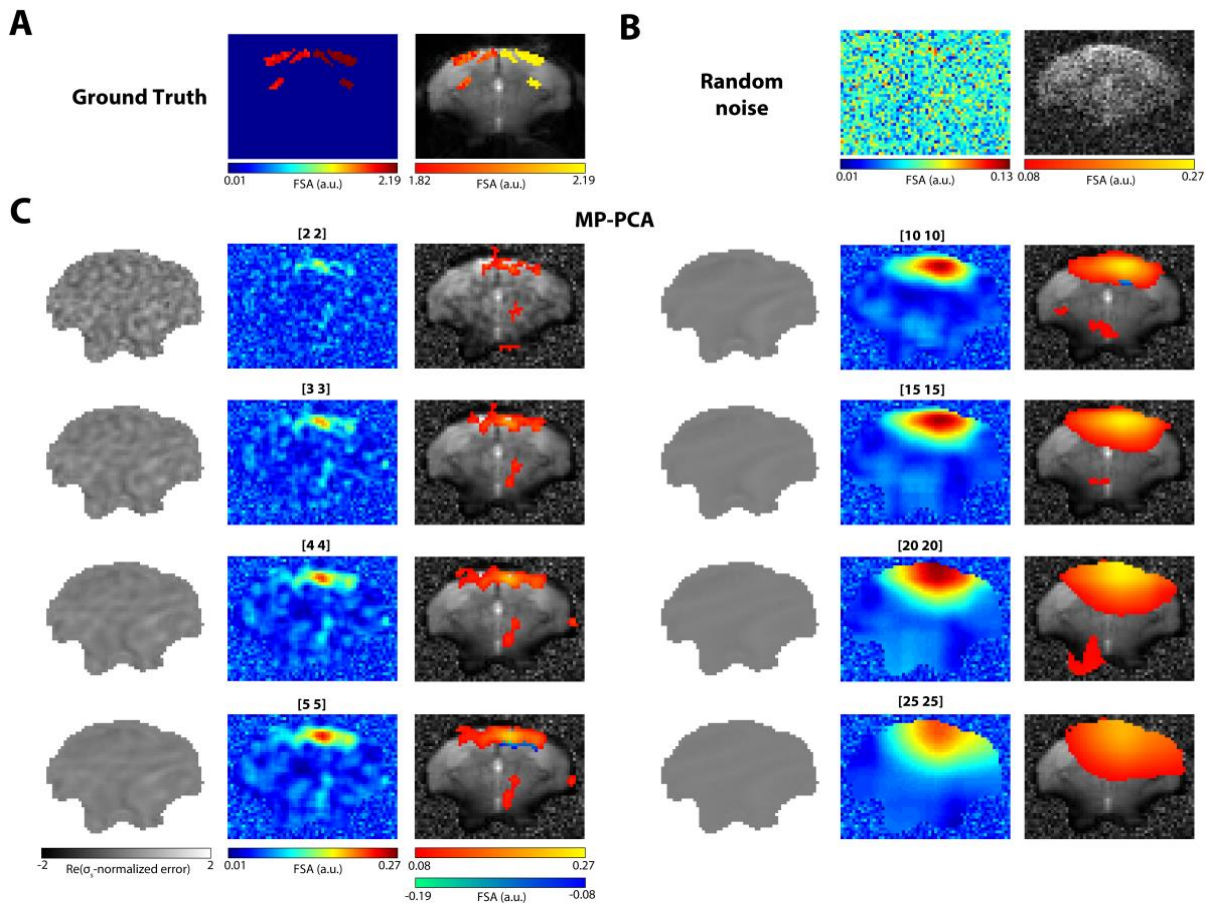

**Figure S10 – Error and BOLD maps of simulated ultrafast fMRI data (with tSNR = 3) after MP-PCA denoising. (A)** BOLD Fourier map of ultrafast fMRI data simulated without any source of noise, with (*right*) and without (*left*) threshold. **(B)** Same as (A) after addition of Gaussian white noise so that the average tSNR in the brain = 3. **(C)** Maps of $\sigma_s$-normalized error of a single frame (*left*) and BOLD Fourier maps with (*right*) and without (*middle*) threshold upon MP-PCA denoising with 8 different sliding windows. Maps are thresholded with a minimum sum of FSA at paradigm's fundamental frequency and two following harmonics = 0.08 and a minimum cluster size = 8. A representative image of the data is shown below the maps.

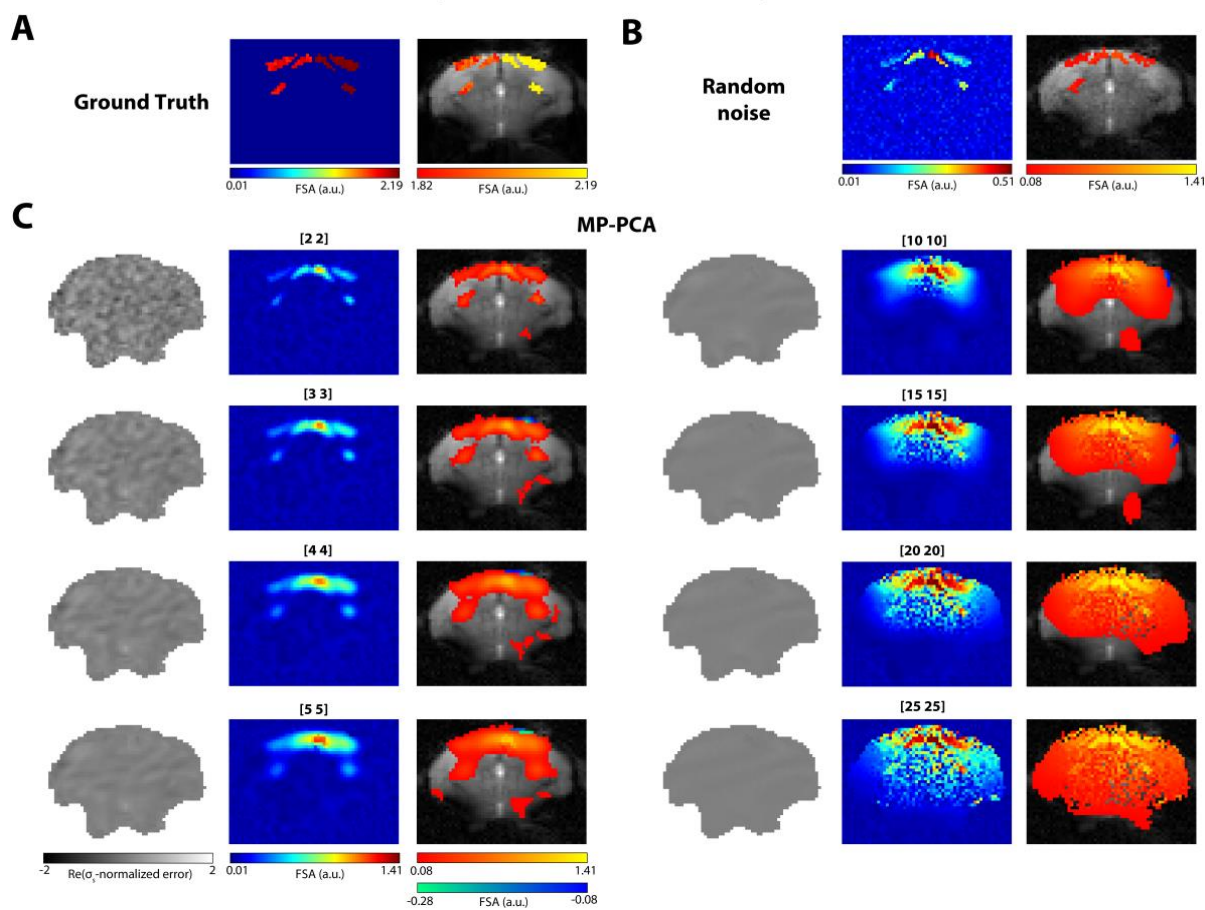

**Figure S11 – Error and BOLD maps of simulated ultrafast fMRI data (with tSNR = 15) after MP-PCA denoising. (A)** BOLD Fourier map of ultrafast fMRI data simulated without any source of noise, with (*right*) and without (*left*) threshold. **(B)** Same as (A) after addition of Gaussian white noise so that the average tSNR in the brain = 15. **(C)** Maps of $\sigma_s$-normalized error of a single frame (*left*) and BOLD Fourier maps with (*right*) and without (*middle*) threshold upon MP-PCA denoising with 8 different sliding windows. Maps are thresholded with a minimum sum of FSA at paradigm's fundamental frequency and two following harmonics = 0.08 and a minimum cluster size = 8. A representative image of the data is shown below the maps.

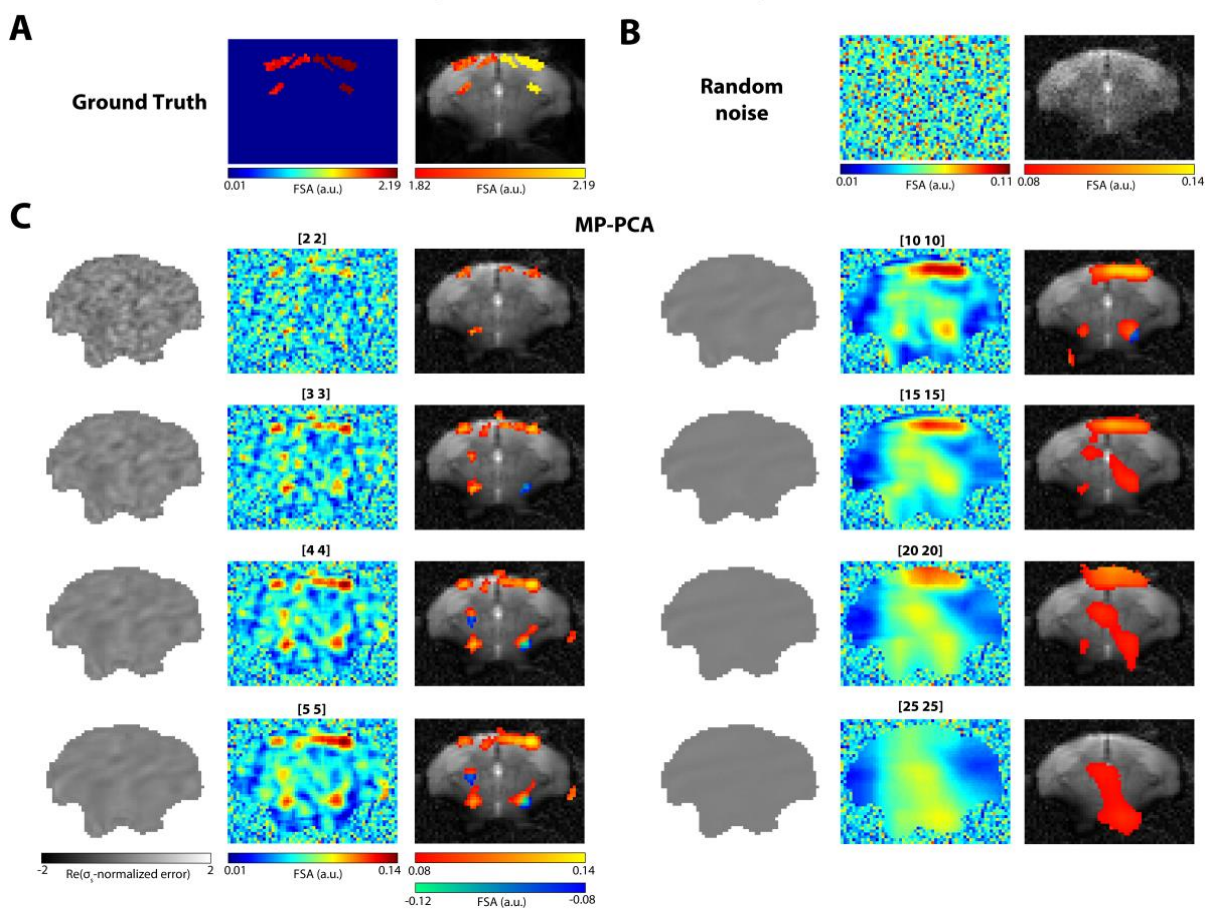

**Figure S12 – Error and BOLD maps of simulated ultrafast fMRI data (with 0.10-0.70% BOLD changes) after MP-PCA denoising.** **(A)** BOLD Fourier map of ultrafast fMRI data (with 0.10-0.70% BOLD changes) simulated without any source of noise, with (*right*) and without (*left*) threshold. **(B)** Same as (A) after addition of Gaussian white noise so that the average tSNR in the brain = 6.7. **(C)** Maps of $\sigma_S$-normalized error of a single frame (*left*) and BOLD Fourier maps with (*right*) and without (*middle*) threshold upon MP-PCA denoising with 8 different sliding windows. Maps are thresholded with a minimum sum of FSA at paradigm's fundamental frequency and two following harmonics = 0.08 and a minimum cluster size = 8. A representative image of the data is shown below the maps.

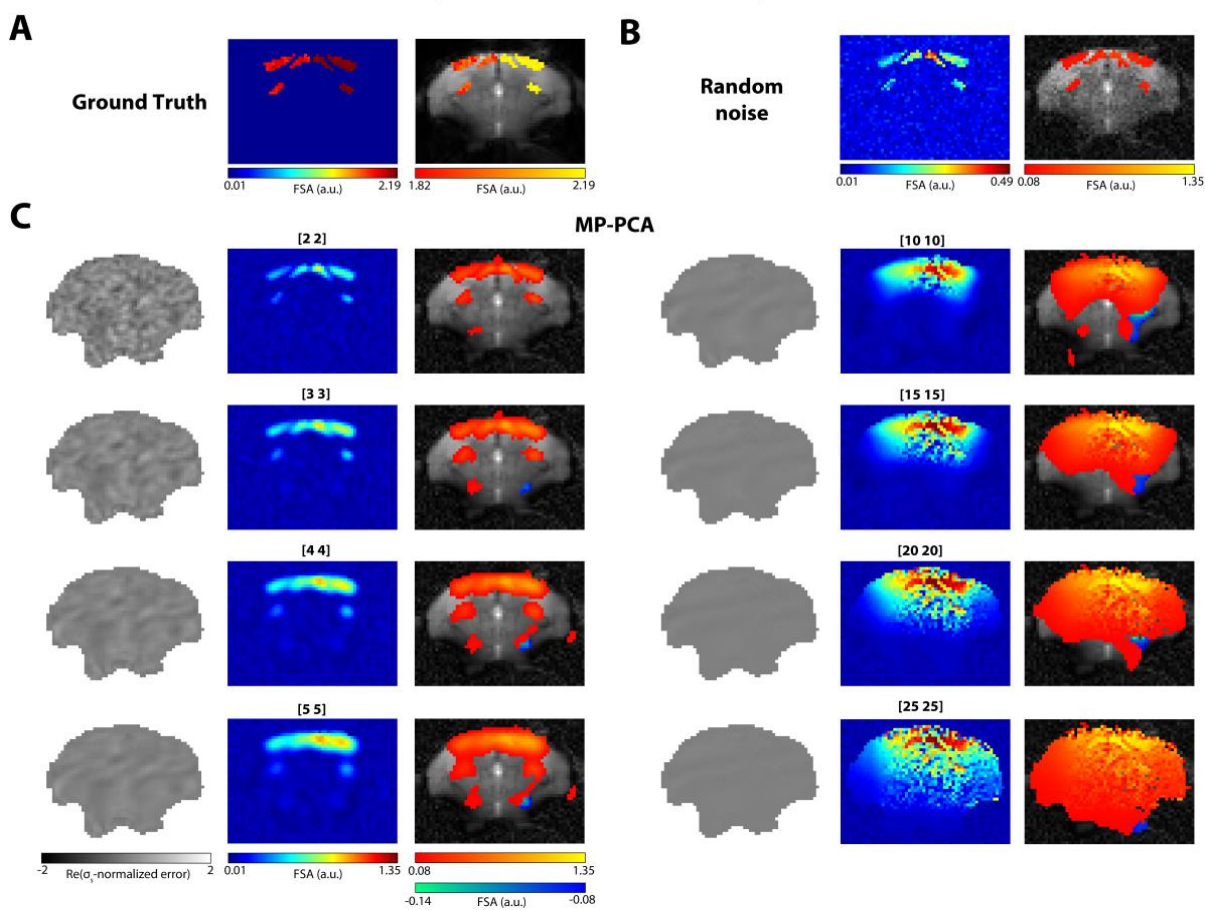

**Figure S13 – Error and BOLD maps of simulated ultrafast fMRI data (with 2.75-4.25% BOLD changes) after MP-PCA denoising. (A)** BOLD Fourier map of ultrafast fMRI data (with 2.75-4.25% BOLD changes) simulated without any source of noise, with (*right*) and without (*left*) threshold. **(B)** Same as (A) after addition of Gaussian white noise so that the average tSNR in the brain = 6.7. **(C)** Maps of $\sigma_S$-normalized error of a single frame (*left*) and BOLD Fourier maps with (*right*) and without (*middle*) threshold upon MP-PCA denoising with 8 different sliding windows. Maps are thresholded with a minimum sum of FSA at paradigm's fundamental frequency and two following harmonics = 0.08 and a minimum cluster size = 8. A representative image of the data is shown below the maps.

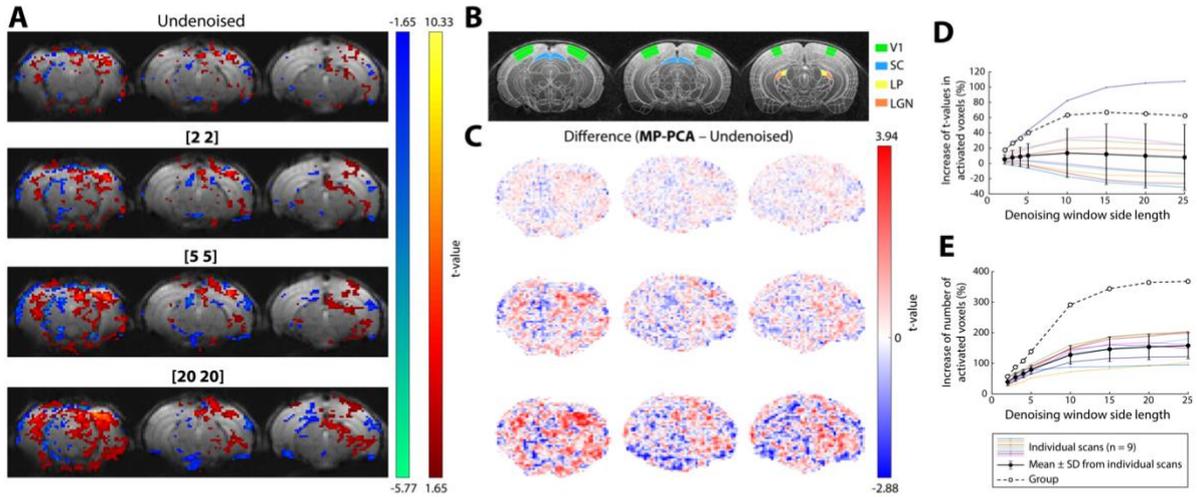

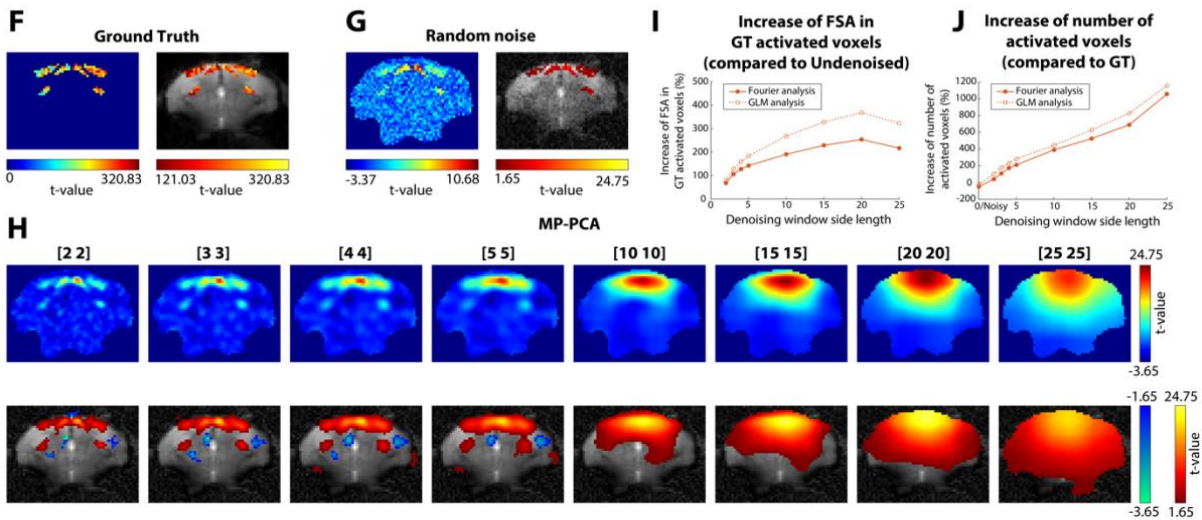

**Figure S14 – Individual GLM BOLD maps of acquired multislice and simulated ultrafast fMRI data after MP-PCA denoising. (A)** BOLD maps of a representative multislice fMRI acquisition (3 out of 10 slices shown) obtained before and after MP-PCA denoising at the slice level with 3 different sliding windows ([2 2], [5 5] and [20 20]) and generated by GLM analysis. ZF was removed from k-space prior to data denoising. Maps are thresholded with a minimum *t*-value = 1.65 and a minimum cluster size = 10. **(B)** ROIs of the mouse visual pathway from the Allen Reference Atlas delineated on anatomical images. **(C)** Difference between the functional maps shown in (A). **(D)** Percentage increase of t-values in activated voxels relative to the undenoised data results across 8 different sliding windows and n = 9 different scans (represented by each color plot) of multislice fMRI data. **(E)** Percentage increase of number of activated voxels in the BOLD maps relative to the undenoised data results across the same sliding windows and scans. The empty dots and dashed lines in (D) and (E) are the results from the group BOLD maps. **(F)** BOLD map of ultrafast fMRI data (with 0.75-2.25% BOLD changes) simulated without any source of noise, with (*bottom*) and without (*top*) threshold, generated by GLM analysis. **(G)** Same as (F) after addition of Gaussian white noise so that the average tSNR in the brain = 6.7. **(H)** Same as (G) after MP-PCA denoising with 8 different sliding windows. A representative image of the data is shown below the maps. Maps are thresholded

with a minimum *t*-value = 1.65 and a minimum cluster size = 8. **(I)** Percentage increase of t-values in GT activated voxels relative to the undenoised data results and **(J)** Percentage increase of number of activated voxels in the BOLD maps relative to the GT data results across 8 different sliding windows in this simulated ultrafast fMRI dataset. Empty dots with dashed lines and filled dots with solid lines represent the variations observed in maps generated using GLM and Fourier analysis, respectively.